\documentclass[11pt,a4paper]{article}

\pdfoutput=1
\usepackage{graphicx}
\usepackage{amssymb}
\usepackage{amsmath}
\usepackage{mathtools}
\usepackage{mathpazo}
\usepackage[mathpazo]{flexisym}
\usepackage{breqn}
\usepackage{bm}
\usepackage{latexsym}
\usepackage{epsfig}
\usepackage{float}
\usepackage{textcomp}
\usepackage{color}
\usepackage{float}
\usepackage{blindtext}
\usepackage{enumitem}
\usepackage{xcolor}
\usepackage[section]{placeins}
\usepackage{caption}
\usepackage{float}
\usepackage{siunitx}
\usepackage{tikz} % To generate the plot from csv
\usepackage{pgfplots}
\usepackage{subcaption}
\usepackage[labelformat=parens,labelsep=quad,skip=3pt]{caption}
\usepackage{graphicx}
\pgfplotsset{compat=newest} % Allows to place the legend below plot
\usepgfplotslibrary{units} % Allows to enter the units nicely

\usepackage{calligra}
\usepackage[colorlinks=true,linkcolor=blue,citecolor=blue,urlcolor=blue,linktocpage=true]{hyperref}
\DeclareMathAlphabet{\mathcalligra}{T1}{calligra}{m}{n}
\DeclareFontShape{T1}{calligra}{m}{n}{<->s*[2.2]callig15}{}

\usepackage[mathscr]{euscript}
\usepackage[export]{adjustbox}
\def\l{\left}
\def\r{\right}

\def\be{\begin{equation}}
\def\ee{\end{equation}} 
\def\bea{\begin{eqnarray}}
\def\eea{\end{eqnarray}}

\newcommand{\abs}[1]{\left|#1\right|}
\def\lsim{\mathrel{\rlap{\lower4pt\hbox{\hskip0.5pt$\sim$}}
 \raise1pt\hbox{$<$}}}         %less than or approx. symbol
\def\gsim{\mathrel{\rlap{\lower4pt\hbox{\hskip0.5pt$\sim$}}
 \raise1pt\hbox{$>$}}}         %greater than or approx. symbol

\newcommand{\dd}{\mathrm{d}}

\newcommand{\vk}{\mathbf{k}}
\newcommand{\vp}{\mathbf{p}}

\newcommand{\T}{\mathcal T}

\newcommand{\Ph}{\mathcal P_h}

\UseRawInputEncoding
\numberwithin{equation}{section}

\begin{document}

\providecommand{\abs}[1]{\lvert#1\rvert}
\providecommand{\bd}[1]{\boldsymbol{#1}}

\begin{titlepage}

\setcounter{page}{1} \baselineskip=15.5pt \thispagestyle{empty}

\begin{flushright}
%% preprint number
\end{flushright}
%\vfil

\bigskip
\begin{center}
{\LARGE \textbf{Small scale induced gravitational waves from primordial }}\\
 \medskip
 {\LARGE \textbf{black holes, a stringent lower mass bound and the }}\\ 
 \medskip
 {\LARGE \textbf{imprints of an early matter to radiation transition}}
\vskip 15pt
\end{center}

\vspace{0.5cm}
\begin{center}
{\Large Nilanjandev Bhaumik and Rajeev Kumar Jain}
\end{center}

\vspace{0.2cm}
\begin{center}
\textit {\large Department of Physics, Indian Institute of Science,\\
Bangalore 560012, India}
\vskip 14pt
E-mail:
 \texttt{\href{mailto:nilanjandev@iisc.ac.in}{nilanjandev@iisc.ac.in}},
 \texttt{\href{mailto:rkjain@iisc.ac.in}{rkjain@iisc.ac.in}}
\end{center}

In all inflationary scenarios of primordial black holes (PBH) formation, amplified scalar perturbations inevitably accompany an induced stochastic gravitational waves background (ISGWB) at smaller scales. 
In this paper, we study the ISGWB originating from the inflationary model, introduced in our previous paper \cite{Bhaumik:2019tvl} wherein PBHs can be produced with a nearly monochromatic mass fraction in the asteroid mass window accounting for the total dark matter in the universe.  We numerically calculate the ISGWB in our scenario for frequencies ranging from nanoHz to KHz that covers the observational scales corresponding to future space based GW observatories such as IPTA, LISA, DECIGO and ET. Interestingly, we find that ultralight PBHs ($M_{\rm PBH} \sim 10^{-20} M_\odot$) which shall completely evaporate by today with exceedingly small contribution to dark matter, would still generate an ISGWB that may be detected by a future design of the ground based Advanced LIGO detector.
Using a model independent approach, we obtain a stringent lower mass limit for ultralight PBHs which would be valid for a large class of ultra slow roll inflationary models. Further, we extend our formalism to study the imprints of a reheating epoch on both the ISGWB and the derived lower mass bound. We find that any non-instantaneous reheating leads to an even stronger lower bound on PBHs mass and an epoch of a prolonged matter dominated reheating shifts the ISGWB spectrum to smaller frequencies. In particular, we show that an epoch of an early matter dominated phase leads to a secondary amplification of ISGWB at much smaller scales corresponding to the smallest comoving scale leaving  the horizon during inflation or the end of inflation scale.  
Finally, we discuss the prospects of the ISGWB detection by the proposed and upcoming GW observatories.

%\end{abstract}
%\vfil
%\keywords{Inflation, Primordial black holes, Gravitational waves, Dark matter}
%\arxivnumber{}
\end{titlepage}

\newpage
\tableofcontents

%%######################################################################################%%

\section{Introduction}
%\subsection{BHs and GWs}

Primordial black holes (PBH)  are now widely considered one of the most interesting  candidate to explain the cold dark matter (CDM) in the universe and have gained a lot of attention lately, thanks to the recent detection of astrophysical gravitational waves (GW) from a system of binary black holes, as reported by the LIGO-Virgo scientific collaboration \cite{Abbott:2016blz, Abbott:2016qqj, Abbott:2016wfe, Abbott:2016nmj, Abbott:2017vtc, Abbott:2017oio}. 
Moreover, it has been discussed that super massive black holes which are observed at the centre of massive galaxies at high redshifts could have been originated from the distribution of PBHs \cite{Abbott:2016htt, Belczynski:2016obo, Duechting:2004dk, Bean:2002kx}. 
It is well known that PBHs can be produced in the early universe, particularly, after inflation when primordial curvature perturbations with large overdensities re-enter the horizon during the radiation dominated (RD) epoch \cite{Hawking:1971ei, Carr:1974nx, Khlopov:1985jw, Ivanov:1994pa}. Lately, a large number of inflationary models have been studied to produce PBHs in different mass ranges, in particular, the class of models producing PBHs in the asteroid mass window in which PBHs could contribute to the total CDM in the universe \cite{GarciaBellido:1996qt, Kawaguchi:2007fz, Kohri:2007qn, Drees:2011yz, Bugaev:2011wy, Erfani:2013iea, Clesse:2015wea, Erfani:2015rqv, Garcia-Bellido:2017mdw, Ezquiaga:2017fvi, Kannike:2017bxn, Ballesteros:2017fsr, Hertzberg:2017dkh, Pi:2017gih, Cicoli:2018asa, Kamenshchik:2018sig, Dimopoulos:2019wew, Atal:2019erb, Mishra:2019pzq, Cai:2019bmk, Cheong:2019vzl, Ballesteros:2020qam, Palma:2020ejf, Conzinu:2020cke, Braglia:2020eai}. 

A stochastic background of primordial GWs is a central prediction of all the inflationary models. In particular, a nearly scale invariant spectrum of tensor perturbations is widely regarded as the holy grail of canonical single field slow roll inflationary models. Such a background encodes pivotal information which can be used to probe and constrain the physics of the early universe and fundamental physics operating at very high energies. In single field slow roll inflationary models, the nearly scale invariant GWs background is usually generated from the amplification of vacuum tensor fluctuations at the linear order wherein the inflaton field is the only dynamical degree of freedom. However, models with many dynamical fields such as extra spectator fields or gauge fields also lead to a secondary background of GWs with very different properties and characteristics than the vacuum contribution. Distinct signatures of such a background can then be used to probe a general class of inflationary scenarios beyond the single field inflationary models. 
Besides probing the fundamental physics of the early universe, the spectral energy density of inflationary GWs at the present epoch can also be used to trace and probe the thermal history of the universe \cite{Starobinsky:1979ty, Boyle:2005se, Durrer:2011bi, Kuroyanagi:2010mm, Lozanov:2016hid, Wang:2016ana, Figueroa:2019paj}. For some recent reviews on various cosmological backgrounds of GWs, their generation and their detection, see refs. \cite{Guzzetti:2016mkm, Caprini:2018mtu}.

In general, an induced stochastic gravitational waves background (ISGWB) at second order through mode coupling of scalar metric perturbations is always generated in all the inflationary models. Since the power spectrum of second order metric perturbations should be of order $10^{-18}$ in the RD era, one can expect that this GWs background to be extremely small and is quite far from the reach of present and upcoming GWs observatories. However, in all the inflationary scenarios of PBHs formation, since the power spectrum of scalar curvature perturbations must be enhanced to $\sim 10^{-2}$ at smaller scales to produce PBHs, this ISGWB typically turns out to be quite large and is therefore, considered an interesting and relevant byproduct of all such inflationary scenarios \cite{Ananda:2006af, Baumann:2007zm, Saito:2008jc, Alabidi:2012ex, Alabidi:2013lya, Nakama:2016gzw, Kohri:2018awv, Cai:2018dig, Bartolo:2018evs, Bartolo:2018rku, Inomata:2018epa, Clesse:2018ogk, Cai:2019amo, Hajkarim:2019nbx, Fu:2019vqc, Lin:2020goi, Domenech:2020kqm, Dalianis:2020cla, Ragavendra:2020sop, Unal:2020mts}.

In an earlier paper \cite{Bhaumik:2019tvl}, we had studied an inflationary scenario with a sextic order polynomial potential that allows the existence of an inflection point in the inflaton potential. Such a potential allows an epoch of an ultra slow roll (USR) evolution which leads to an enhancement of the spectrum of primordial scalar curvature perturbations at small scales. We found that this scenario can produce PBHs in different mass ranges and in particular, in the asteroid mass range in which PBHs can contribute to the entire CDM. 
In this paper, we shall study the ISGWB arising in this scenario which is generated from the contribution due to mode coupling of first order scalar perturbations at quadratic order.
%\footnote{Although we are using the phrase {\it second order scalar perturbations}, it evidently refers to the first order scalar perturbations at quadratic order in the source term and not the intrinsic second order scalar perturbations.}   
Such GWs are generated on smaller scales after they re-enter the horizon during the RD phase. We shall calculate this ISGWB in our model which is produced in a range of different frequencies from nano-Hz to KHz, using an adequate numerical approach and compare it with the current and projected sensitivities of various ground and future space based GWs observatories. Interestingly, we find that in our model, this ISGWB can be simultaneously detected by different GWs observatories which is usually not the case when the primordial power spectrum is highly peaked. We emphasise that this feature provides a unique opportunity to constrain the resulting signal much better. In models which produce PBHs in the asteroid mass window as in our scenario, the secondary GWs background is usually peaked in the frequency band $f_{{\rm GW}} \sim 10^{-3} - 1\, {\rm Hz}$ and thus, can be potentially detected by the future space based GWs observatories such as LISA \cite{Danzmann:1997hm, Sathyaprakash:2009xs, Bartolo:2016ami, Audley:2017drz}, TAIJI \cite{Guo:2018npi}, DECIGO \cite{Seto:2001qf, Yagi:2011wg} or BBO \cite{Corbin:2005ny}.
Moreover, we notice an interesting observational possibility that ultralight PBHs produced in our scenario with mass $M_{\rm PBH} \sim 10^{-20} - 10^{-21} \, M_\odot$  which would have been completely evaporated by today, would still lead to an ISGWB at larger frequencies which can, in principle, be observed by the future design of the ground based Advanced LIGO detector \cite{ALIGO:2014jea}.

Furthermore, using a model independent approach, we obtain a robust lower bound on the PBHs mass in our case by assuming an instantaneous as well as a smooth transition from the USR to the slow roll (SR) phase. Our mass bounds are applicable as far as there is no intermediate fast roll stage between the USR and the SR phase. We then extend our formalism to study the imprints of a reheating epoch on both the ISGWB and the derived lower mass bound. We find that a prolonged epoch of a non instantaneous reheating leads to a shift in the ISGWB energy spectrum towards smaller frequencies. Thus, such ISGWB signal can be detected simultaneously by different GWs observatories. Finally, we study the imprints of a transition from an early matter dominated (eMD) phase to a RD phase on the ISGWB and find that such a transition leads to a secondary amplification of ISGWB at much smaller scales corresponding to the smallest comoving scale leaving the horizon during inflation or the end of inflation scale.

The remainder of this paper is organised as follows: In the following section, we shall quickly discuss the basic formalism to compute the ISGWB from first order scalar perturbations in any general inflationary scenario.  We shall compute the spectral energy density of GWs today for the scenario of PBHs formation of our earlier work and compare it with the optimal (design) sensitivities of various present and future GWs observatories. We shall show that the ISGWB induced by ultralight PBHs can be detected by a future run of the ground based Advanced LIGO detector. In Section \ref{low-mass-pbh}, we shall discuss in detail how to obtain the lowest possible PBHs mass both for the cases of an instantaneous transition and a smooth transition from the USR to the SR phase. In Section \ref{reh-gw}, we shall study the resulting effects of a non instantaneous reheating epoch on the GWs spectra and the lower bound on the PBHs mass as well as a secondary amplification of ISGWB due to a transition from an eMD phase to RD phase. Finally, we shall summarise our results and discuss their implications in Section \ref{discuss}. In the appendix \ref{appendix}, we shall provide the details of the calculations of the transition from an eMD to RD era. 

Our conventions and notations adopted in this paper are as follows. We work in the natural units, $\hbar =c=1$, with reduced Planck mass $M_{\rm Pl}^2 = (8 \pi G)^{-1}$. 
%Our metric signature is mostly plus with $(-, +, +, +)$. The background metric is described by the homogeneous, isotropic and spatially flat FLRW universe with a line element $ds^2 = -dt^2 +a^2(t) d{\bf x}^2 = a^2(\tau) (-d\tau^2 + d{\bf x}^2)$. 
The conformal time $\tau$ is defined as $d\tau =dt/a(t)$. The overdots and primes denote the derivatives with respect to the cosmic time $t$ and the conformal time $\tau$, respectively. The Hubble parameter is defined as $H \equiv {\dot a}/a$ while the conformal Hubble parameter is given by ${\cal H} \equiv a H \equiv {a'}/a$.
 
%%%%%%%%%%%%%%%%%%%%%%%%%%%%%%%%%%%%%%%%%%%%%%%%%%%%%%%%%%%%%%%%%%%%%%

\section{Stochastic GWs from first order scalar perturbations}
\label{secondarygw}

It is well known that, at the linear order in perturbations, the scalar, vector and tensor perturbations evolve independently, thanks to the decomposition theorem and their evolution is governed by their corresponding equations of motion. 
%In particular, the vector perturbations simply decay in an inflating universe. 
However, at the second order in perturbations, an extra source term is generated for the tensor perturbations due to the mode coupling of scalar metric fluctuations which inevitably leads to an ISGWB. In this section, we shall present the basic formalism and essential equations for the tensor perturbations with a source term due to first order scalar perturbations at quadratic order. We shall then solve these equations numerically for the inflationary scenario of our earlier work \cite{Bhaumik:2019tvl} and calculate the GWs energy density spectrum at the present epoch. We shall also discuss the potential detection of this ISGWB with the future space based GWs observatories such as IPTA, LISA, DECIGO and ET.

In this section, we shall closely follow the discussion of the seminal paper \cite{Ananda:2006af}. Let us start with perturbing the FLRW metric with the scalar and tensor perturbations. In the conformal Newtonian gauge, the perturbed metric can be written as 
\begin{equation}
ds^2 = -a^2(\tau)(1+2\Phi)\, d\tau^2 + a^2(\tau) \left[(1-2\Psi)\delta_{ij}+\frac{1}{2}h_{ij} \right] dx^i dx^j,
\end{equation}
where $\Phi$ and $\Psi$ are the scalar metric perturbations, also called the Bardeen potentials and $h_{ij}$ is the tensor perturbation 
which is symmetric $(h_{ij} = h_{ji})$, traceless ($h_{ii} = 0)$, and transverse $(h_{ij,j} = 0)$. Assuming $\Phi = \Psi$, the scalar part of the anisotropic stress (shear) vanishes but the corresponding tensor part does not. As we shall see later, this serves as a source to the evolution equation for tensor perturbations. With the Fourier modes $h_{{\bf k}}$, the dimensionless power spectrum $\mathcal{P}_h$ is
\begin{eqnarray}
\frac{k^3}{2 \pi^2}\l\langle h_{{\bf k}}^{\lambda} (\tau) \, h_{{\bf k}'}^{\lambda'} (\tau) \r\rangle = \delta_{\lambda \lambda'} \delta^3 ({\bf k} + {\bf k}')  \mathcal{P}_h (\tau, k),
\end{eqnarray}
where $\lambda, \lambda' = \{+,\times\}$ represent the two polarisations of tensor perturbations. Now, the GWs energy density per logarithmic wavelength can be defined as
\begin{equation}
\Omega_{\text{GW}}(\tau, k) \equiv \frac{1}{\rho_c}\frac{\dd\, \rho_{\text{GW}}}{\dd\, {\rm ln} k}=\frac{\rho_{\text{GW}}(\tau,k)}{\rho_{\text{tot}}(\tau)}= \frac{1}{24} \left( \frac{k}{{\cal H}} \right)^2 \overline{\mathcal{P}_h(\tau, k)}, \label{Omega_1}
\end{equation}
where the overline denotes an average over time. The observationally relevant quantity is the energy density $\Omega_{\text{GW}}$ at the present epoch $\tau = \tau_{0}$. Note that, in parity invariant scenarios as in our model, both the polarisations  will lead to the same result for the GWs spectrum. However, in parity violating situations, the power spectrum will be different for the two polarisations. In particular, when one helicity mode is exponentially amplified due to dynamical instabilities than the other, the power spectrum turns out to be maximally helical and has very interesting observational implications. For simplicity, from now on, we shall ignore the superscript $\lambda$ in $h_{\bf k}$.  

\subsection{Induced tensor modes and their power spectrum}

Using the standard canonical quantisation procedure for $h_{ij}$, one finds that the equation of motion for the Fourier modes $h_{\bf k}$, sourced by the scalar perturbations $\Phi$ is given by
\begin{equation}
h_{\bf k}''(\tau) + 2 \mathcal{H} h'_{\bf k}(\tau)+ k^2 h_{\bf k}(\tau) = 4 S_{\bf k}(\tau),  \label{EOM_h}
\end{equation}
where $S_{\bf k}$ is the Fourier component of the source term comprising of first order scalar perturbations. 
This differential equation can be solved by the Green's function method which yields the solution as \cite{Kohri:2018awv}
\begin{equation}
h_{\bf k}(\tau) = \frac{4}{a(\tau)} \int^\tau \text{d}\bar{\tau}\, a(\bar{\tau}) \,G_{\bf k}(\tau, \bar{\tau})\,  S_{\bf k}(\bar{\tau}), 
\label{h_sol}
\end{equation}
where $G_{\bf k}(\tau, \bar{\tau})$ is the solution to the following equation
\begin{equation}
G_{\bf k}''(\tau, \bar{\tau}) +\left[ k^2 - \frac{ a''(\tau)}{a(\tau)}\right] G_{\bf k}(\tau, \bar{\tau}) = \delta (\tau - \bar{\tau}). \label{EOM_Green}
\end{equation}
Since we are interested in the ISGWB on smaller scales corresponding to $k \gg  k_{\rm eq}$ which re-enter the horizon during the RD epoch, we shall restrict our following discussion to $w=1/3$ only. In RD universe $(a \sim \tau$), we can express $G_{\bf k}(\tau, \bar{\tau})$ as
\bea
G_{\bf k}(\tau, \bar{\tau})=\frac{1}{k}\sin\,(x-\bar{x}),
\eea
where $x=k\tau$ and $\bar{x}=k\bar{\tau}$, respectively. While $G_{\bf k}(\tau, \bar{\tau})$ involves the effects of propagation for GW wave, the effects of amplified scalar perturbation comes from the source term, and it depends on the time evolution of the scalar perturbation modes.
During the RD era, ISGWB is produced mainly around the horizon re-entry, without growing any further because the gravitational potential oscillates after horizon re-entry. 
At first order, the time evolution of  $\Phi_{\bf k}$ in RD is governed by
\begin{equation}
\Phi''_{\bf k}(\tau) + \frac{4}{\tau } \Phi'_{\bf k}(\tau) + \frac{1}{3} k^2 \Phi_{\bf k}(\tau)=0. \label{EOM_Phi}
\end{equation}
We can split $\Phi_{\bf k}(\tau)$ into a primordial part, $\tilde \Phi_{\vk}$ (the value at the start of RD) and the transfer function $\T(k\tau)$, representing the time evolution as
%\bea
$\Phi_{\bf k}(\tau) = \T(k\tau) \, \tilde \Phi_{\bf k}$.
%\eea
Note that, in the RD era, the scalar perturbation $\Phi_{\vk}$ is directly related to the gauge invariant comoving curvature perturbation by $\tilde \Phi_{\vk} = \tfrac 23 {\cal R}({\vk})$. The full solution for equation (\ref{EOM_Phi}) for $\T(k\tau)$ can be found in appendix \ref{appendix}. For an instantaneous reheating history, we assume, for $\tau \to 0$, $\T(k\tau) \to 1$ and  $\T'(k\tau) \to 0$, and we can express $\T(k \tau)$ as a function of $x$ as
\begin{equation}
\T(x)= \frac{9}{x^2}\left[ \frac{\sqrt{3}}{x} \sin \l(\frac{x}{\sqrt{3}}\r) -\cos\l(\frac{x}{\sqrt{3}}\r) \right]. 
\label{eq: transfer}
\end{equation}
One can find the detailed calculation of the induced tensor spectrum in \cite{Kohri:2018awv, Bartolo:2018rku, Espinosa:2018eve} and thus, we shall directly write the final expression of the second order tensor power spectra.   

The calculation of the tensor power spectrum ${\cal P}_h$ will involve the four point functions of ${\cal R}(\vk)$. However, we can assume it to be Gaussian at leading order and use Wick's theorem, to write the four point functions in terms of possible combinations of the two point functions or the power spectrum ${\cal P}_{\cal R}(k)$. After a lot of simplification, one finds
\begin{equation}
\Ph(\tau,k) = 4 \int_{0}^{\infty}\dd v \int_{|1-v|}^{1+v} \dd u
\l(\frac{4 v^2-(1+v^2-u^2)^2}{4 u v}\r)^2 I_{\rm RD}^2(v,u,x) {\cal P}_{\cal R}(kv){\cal P}_{\cal R}(ku),
\label{P-h}
\end{equation}
where $v=p/k, \, u = {|\vk-\vp|}/{k}$ and the factor $I_{\rm RD}$ is defined as
 \begin{equation}{\label{IIR}}
\begin{aligned}
I_{\rm RD}(u,v,x)= & \int_{x_r}^x d\bar{x} \frac{\bar{x}}{x}f(u,v,\bar{x},x_r) k G(\bar{x},x),
   \end{aligned}
\end{equation}
where $x_r=k\tau_r$ corresponds to the conformal time $\tau_r$ at the beginning of RD phase, and $x=k\tau$ corresponds to some late time $\tau$ during RD epoch. For instantaneous reheating, we can take $x_r \to 0$. The factor $I(u,v,x)$ is a very involved function and we have derived its general form for a RD universe preceded by an eMD era in appendix \ref{appendix}, from which the results for instantaneous reheating case can be recovered by taking the appropriate limit $x_r \to 0 $.

To calculate the present energy density of ISGWB, we can evolve the integration $I(u,v,x)$ upto a time when SM degrees of freedom become nonrelativistic and find $\Omega_{GW}(\tau,k)$. Using the entropy conservation, we can then express the present ISGWB energy density $\Omega_{GW}(\tau_0,k)$, in terms of present radiation energy density $\Omega_{r,0}$,  $\Omega_{GW}(\tau,k)$ and a constant $c_g$ as \cite{Espinosa:2018eve}
\bea
\Omega_{GW}(\tau_0,k) =  c_g\, \Omega_{r,0}\,  \Omega_{GW}(\tau,k) =  \frac{1}{24} \left( \frac{k}{{\cal H}} \right)^2 c_g\, \Omega_{r,0}\, \overline{\mathcal{P}_h(\tau, k)},
\eea
where  $c_g \approx 0.4$ if we take the number of relativistic degrees of freedom to be $\sim 106$. In RD, ${\cal H}=a H=1/\tau$, and we can write $k/{\cal H}=k\tau=x$ so taking $k/{\cal H}$ factor inside the power spectra integral and defining ${\cal I}$ with $I={\cal I}/x$, we can write
\bea
\Omega_{GW}(\tau_0,k) = \frac{1}{6} c_g \,\Omega_{r,0}  \int_{0}^{\infty}\dd v \int_{|1-v|}^{1+v} \dd u
\l(\frac{4 v^2-(1+v^2-u^2)^2}{4 u v}\r)^2  \overline {\cal I}_{\rm RD}^2(v,u,x) {\cal P}_{\cal R}(kv){\cal P}_{\cal R}(ku). 
\label{omega-in-u-v}
\eea
In the late time limit $x \to \infty$, for a pure RD universe, one gets
\bea
\!\!\!\!\!\!\!\!\!\!\!\!
\overline {\cal I}_{\rm RD}^2(v,u,x \to \infty) 
=  \frac{1}{2}\l(\frac{3 (u^2+v^2-3)}{4u^3v^3}\r)^2
\biggl[\l( -4uv +(u^2+v^2-3)\,{\rm log}\l|\frac{3-(u+v)^2}{3-(u-v)^2}\r|\r)^2 \biggr. \nonumber \\
 \biggl. \!\! +\pi^2 (u^2+v^2-3)^2 \Theta(u+v-\sqrt{3})\biggr].
 \label{ird}
%\end{align}
\eea
Note that, while this formula is valid for an instantaneous reheating history, a more general expression for $\overline {\cal I}_{\rm RD}^2$, assuming RD phase preceded by an eMD phase is derived in appendix \ref{appendix}.

\begin{figure}[t]
\begin{center}
\includegraphics[width=7.7cm, height=6.0cm]{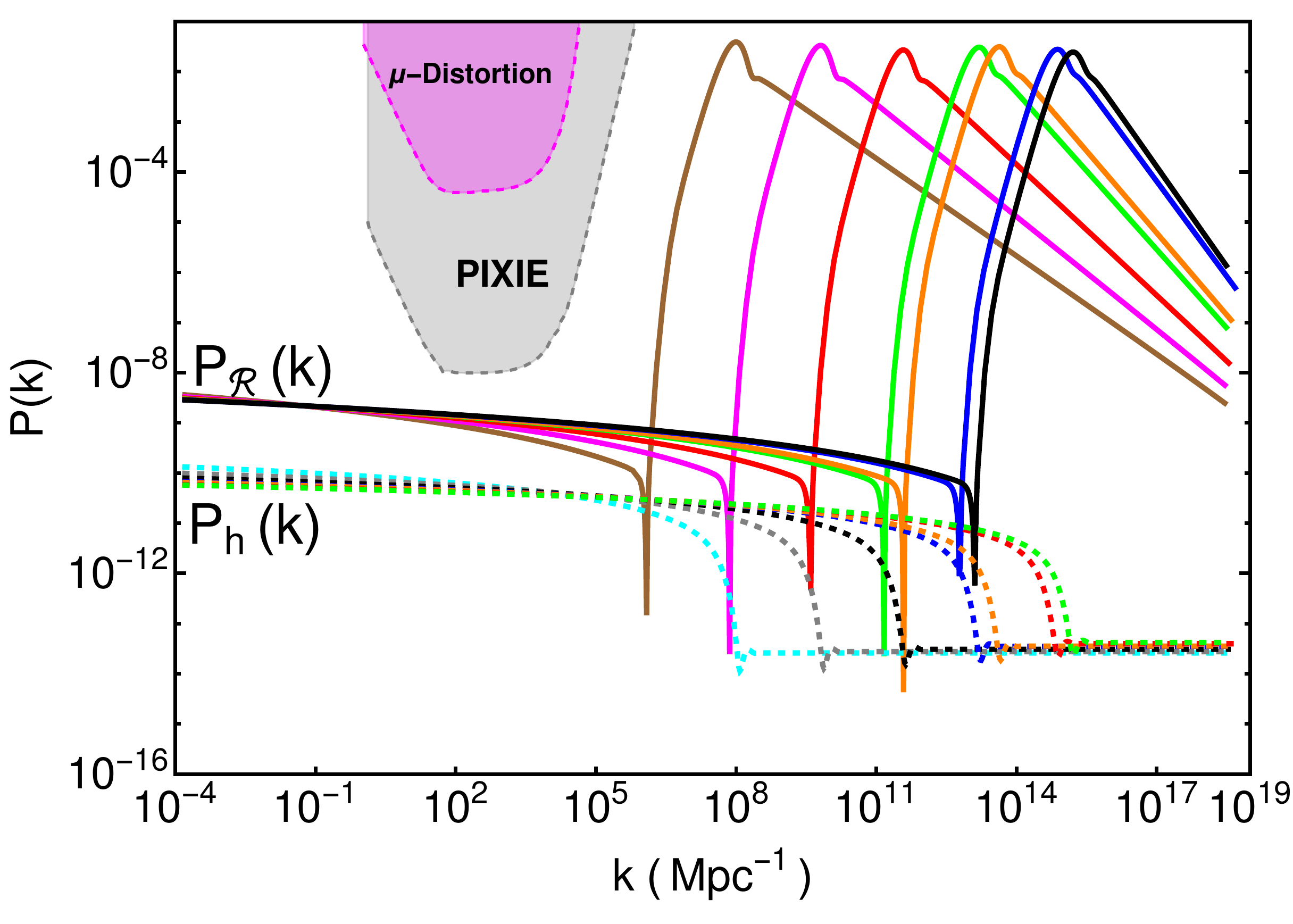}
\hspace{6pt}
\includegraphics[width=7.75cm, height=6.0cm]{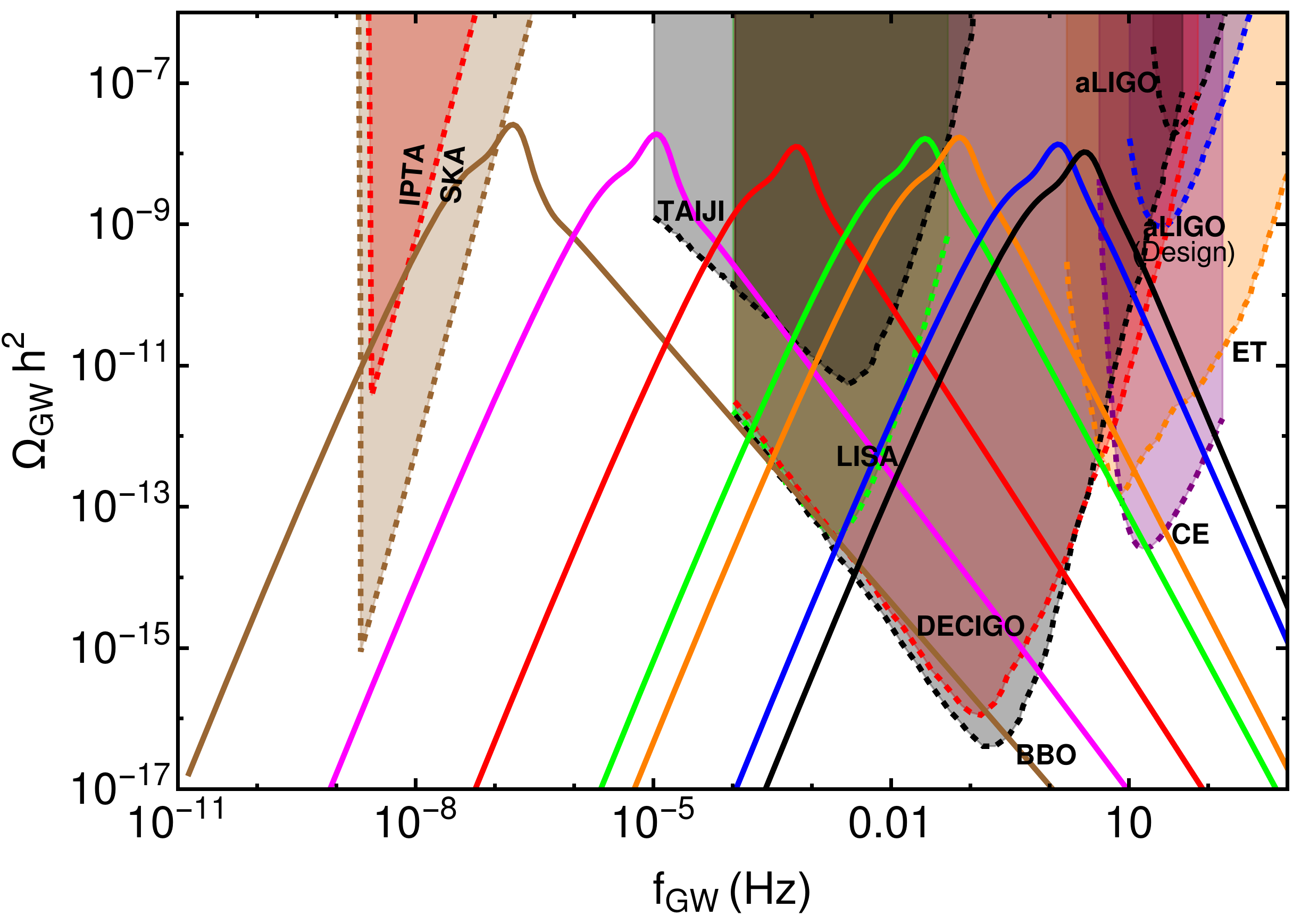}
\vskip 10pt
\caption{On the left, we plot the power spectra of primordial curvature perturbations ${\cal P}_{\cal R}$ (solid curves) and primordial tensor perturbations ${\cal P}_{h}$ (dashed curves) for different parameters of the scenario proposed in our earlier work \cite{Bhaumik:2019tvl}.  All these spectra show a similar enhancement at smaller scales, required for the abundant PBHs production. Also, shown are the relevant constraints from CMB spectral distortions \cite{Chluba:2012we, Chluba:2019nxa}. On the right, we plot the spectral energy density of ISGWB corresponding to the spectra on the left. All the GWs spectra also show a similar behaviour. In particular, a bump in ${\cal P}_{\cal R}$ on small scales leads to a peak in $\Omega_{\text{GW}} h^2$ on larger frequencies which fall in the sensitivity regimes  of various future space based GWs observatories such as IPTA, LISA, DECIGO and ET. The colour coding of different plots is consistent across the two figures.}
\label{sps-gw}
\end{center}
\end{figure}

In our previous work \cite{Bhaumik:2019tvl}, we had developed a numerical code to compute the PBHs mass fraction for inflationary models which allows violations of SR condition, needed for the enhancement of the power spectrum. We have chosen different set of parameters of our model in such a way that the largest possible mass fraction of PBHs is obtained as allowed by various constraints in a given mass range. We have now extended that code by including a routine to compute the ISGWB in such models. In figure \ref{sps-gw} (on the left), we have plotted the power spectra of primordial curvature and tensor perturbations ${\cal P}_{\cal R}$ and ${\cal P}_{h}$  for the scenario that we had discussed in our earlier work \cite{Bhaumik:2019tvl}. The power spectra ${\cal P}_{\cal R}$  correspond to different choices of parameters of the model which leads to different values of the spectral index $n_{_S}$ at the pivot scale. All these spectra show a similar enhancement ${\cal P}_{\cal R} \sim k^4$ at smaller scales, an interesting behaviour which has also been obtained using an analytical formalism \cite{Byrnes:2018txb, Carrilho:2019oqg}. The spectral distortions constraints on the primordial power spectrum derived from COBE/FIRAS and forecasts for PIXIE are also shown \cite{Chluba:2012we}. 
It is well known that the amplitude of the power spectrum of curvature perturbations should be $P_{\cal R} \sim 10^{-2}$ to form PBHs in radiation domination. Assuming a specific functional form of the scalar spectrum, one can arrive at a rough order of magnitude estimation of the PBH non-detection constraints, as shown in \cite{Byrnes:2018txb}. However, this estimate also depends upon the shape of the power spectrum, as has already been shown in \cite{Germani:2018jgr}. It further depends on the choice of the critical density contrast, window function, collapse formalism etc. as discussed in appendix B of our earlier paper \cite{Bhaumik:2019tvl}. Unless one obtains these PBH constrains for the specific model of interest, it is not appropriate to use them for comparison with another model. For this reason, we are not displaying such constraints in our plots in figure \ref{sps-gw}. Moreover, to facilitate easy comparison, we are following the same color coding as in figure 3 of our previous paper \cite{Bhaumik:2019tvl}.

In the right panel of figure \ref{sps-gw}, we have plotted the corresponding spectral energy density $\Omega_{\text{GW}} h^2$ of induced GWs, obtained by numerically integrating equation (\ref{omega-in-u-v}).  We found that all the GWs spectra show a similar behaviour, with a peak at a characteristic frequency determined by the relation (\ref{m-k-f-relation}).
As expected, a bump in ${\cal P}_{\cal R}$ leads to a peak in $\Omega_{\text{GW}} h^2$ which fall in the sensitivity regimes  of various future space based GWs observatories such as IPTA, LISA, DECIGO and ET. As we have found in our scenario, a wider power spectrum ${\cal P}_{\cal R}$ will result in a wider induced GWs spectra spanning a broader frequency range. This is interesting because the induced signal overlaps with the design sensitivity plots of different upcoming GWs observatories, particularly around the mHz - Hz range.  In such a situation, there lies an interesting possibility to simultaneously detect these signals with different observatories and obtain stronger constraints on its origins in terms of the model parameters. It is also interesting to note that, for some cases in our model wherein the bump in the scalar power spectrum is located on rather larger scales (but still much smaller than CMB scales), the resulting induced GWs background can also be detected by an array of future IPTA/SKA detectors \cite{Lentati:2015qwp, Shannon:2015ect}.
Future observations of CMB spectral distortions will also strongly constrain the primordial power spectrum in the regime $1\, {\rm Mpc^{-1}}  \lesssim k \lesssim 10^5\, {\rm Mpc^{-1}}$ \cite{Kogut:2019vqh, Chluba:2019nxa}.

%%%%%%%%%%%%%%%%%%%%%%%%%%%%%%%%%%%%%%%%%%%%%%%%%%%%%%%%

\subsection{Observing ultralight PBHs with Advanced LIGO}
\label{aligo}

As we had mentioned in our earlier work \cite{Bhaumik:2019tvl},  our scenario can produce PBHs in very different mass ranges and all these mass windows are constrained by a variety of observations. However, it turns out that there do not seem to be any observational constraints around the asteroid mass window and thus, PBHs could contribute to the total energy density of CDM around that window, as has been emphasised in the literature recently.  
It is well known that PBHs do evaporate due to Hawking radiation and the evaporation time scale is given by
\bea
t_{\rm ev}(M) \sim \frac{G^2 M^3}{\hbar\, c^4} \sim 10^{63} \l(\frac{M}{M_{\odot}}\r)^3 {\rm yr}.
\label{hawking-evap}
\eea
This implies that PBHs with mass $M \lesssim 10^{-18} M_{\odot}$ ($M \lesssim 10^{15}\, {\rm g}$) would be completely evaporated by today and thus can not contribute to the present density of the CDM in the universe \cite{Carr:2009jm, Carr:2020xqk}. PBHs in the mass range $10^{-18} - 10^{-16} M_{\odot}$ would actually be evaporating at the present epoch and thus can induce an observable $\gamma$-ray background \cite{Laha:2019ssq}. 
However, PBHs in the very low mass range would not contribute to the CDM at all and  would also be completely evaporated by today. However, they might still induce a secondary GWs background which could, in principle, be detected by the future designs of the ground based GWs observatories. 

It is interesting to note that, the three ``peaks" {\it i.e.} the position of the peak in the power spectrum of curvature perturbations, the peak height in the PBHs mass distribution and the frequency of the peak of the GWs signal are related by \cite{Nakama:2016gzw, Garcia-Bellido:2017aan, Bartolo:2018evs}
\bea
\l(\frac{M_{\rm PBH}}{10^{17}\, {\rm g}}\r)^{-1/2} \simeq \frac{k}{2 \times 10^{14}\, {\rm Mpc}^{-1}} = \frac{f}{0.3\, {\rm Hz}},
\label{m-k-f-relation}
\eea
which provides a qualitative understanding of the relation among $M_{\rm PBH}$, $k$ and $f$. This relation roughly indicates that a peak in the power spectrum of curvature perturbations at $k \simeq 2 \times 10^{14}\, {\rm Mpc}^{-1}$ would generate a peak in the GWs spectrum at frequency $f \sim 0.3 \, {\rm Hz}$.  
Moreover, as the sensitivity is maximum for LISA at $f \sim {\rm mHz}$, the peak in ${\cal P}_{\cal R}$ should be around $k \sim 10^{12}\, {\rm Mpc}^{-1}$ which is consistent with what is shown in figure \ref{sps-gw}. This scaling can further be used to roughly figure out what mass range of PBHs can possibly be probed by means of their secondary GWs signatures using the ground based detectors such as Advanced LIGO. The maximal sensitivity of the projected design of the Advanced LIGO detector corresponds to $ f \sim 30 \, {\rm Hz}$. A stochastic GWs signal around this frequency would correspond to very light PBHs with mass around $M_{\rm PBH} \sim 10^{13}\, {\rm g} \sim 10^{-20} M_\odot$. Evidently, from equation (\ref{hawking-evap}), all such PBHs would be completely evaporated through the emission of Hawking radiation from their formation to today and thus, can not constitute the observed abundance of CDM. 

\begin{figure}[t]
\begin{center}
%\vskip -20pt
\includegraphics[width=6.1cm, height=7.8cm]{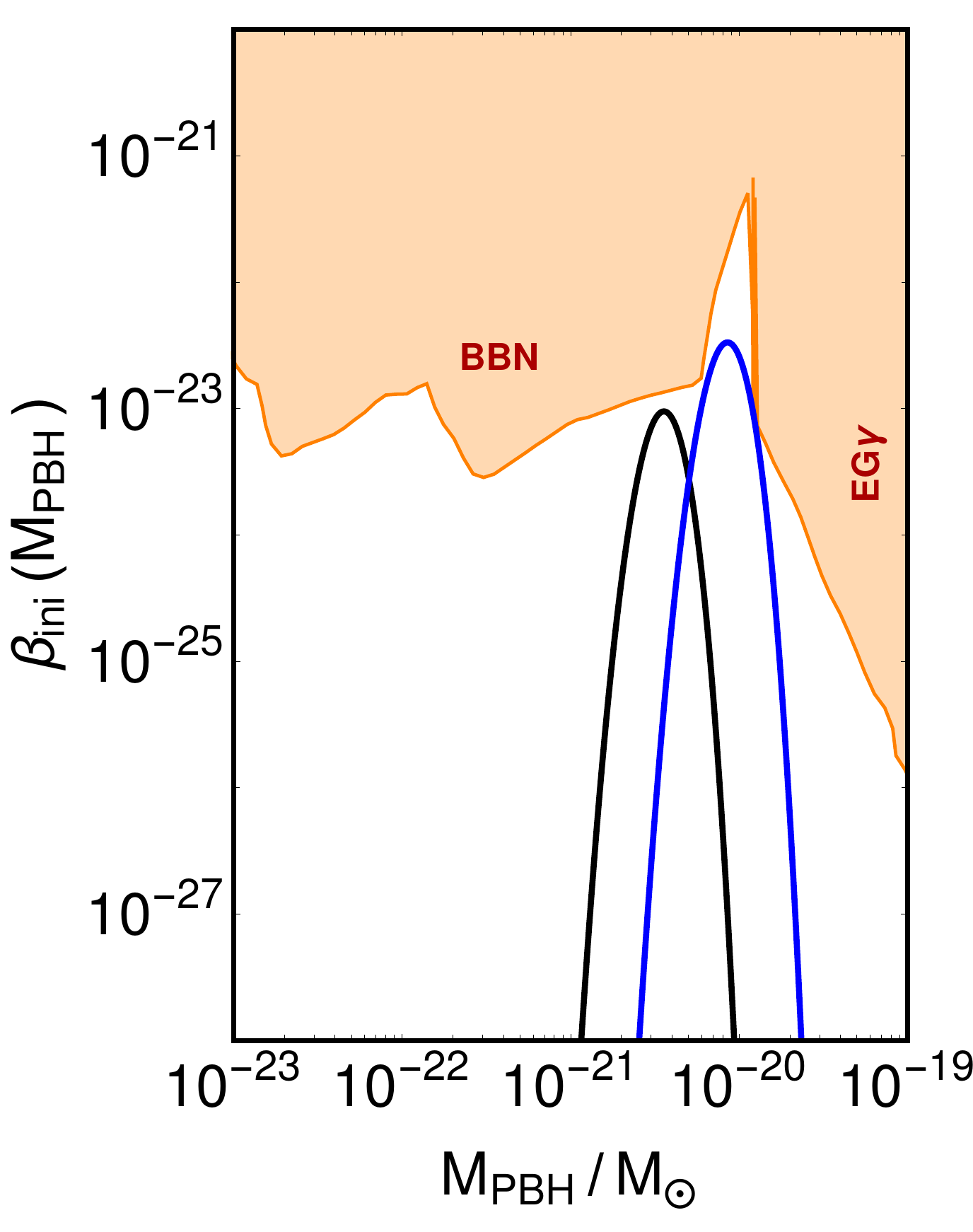}
\hspace{10pt}
\includegraphics[width=6cm, height=7.8cm]{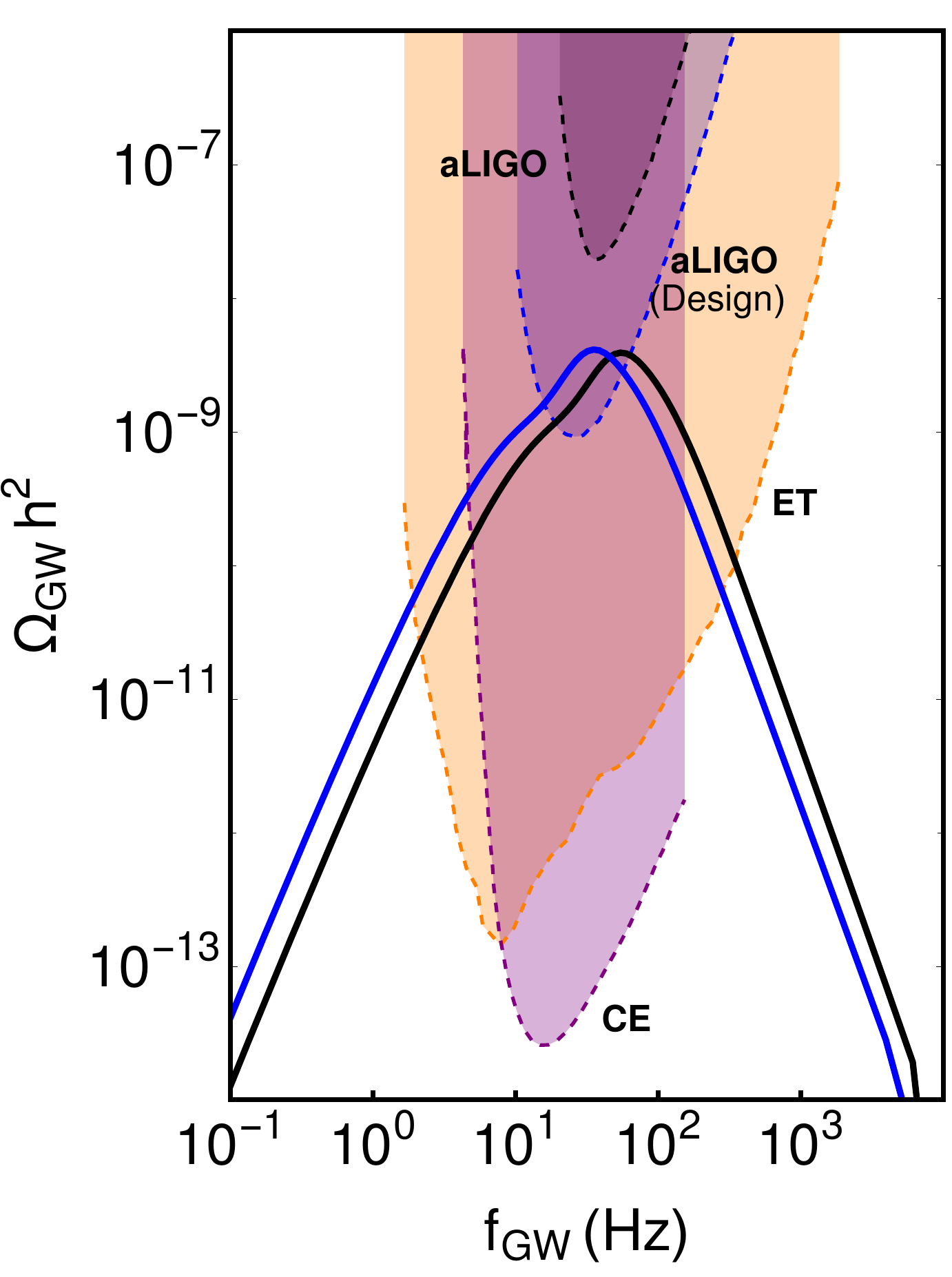}
\vskip 10pt
\caption{On the left, we plot the initial PBHs mass fraction $\beta_{\rm ini}$ (at formation) for two different cases,  produced in our model \cite{Bhaumik:2019tvl} in the very low mass range corresponding to $M \sim 10^{-20} - 10^{-21} M_\odot$, as well as the observational constraints arising from big bang nucleosynthesis (BBN) and extragalactic $\gamma$-ray background. Note that, while comparing with the constraints on $\beta_{\rm ini}$, we are assuming the mass fractions of our model to be monochromatic. Such small mass PBHs would have been completely evaporated by today due to Hawking radiation. However, they will still induce an observable ISGWB in the higher frequency range which falls in the design sensitivity contours of the Advanced LIGO detector and therefore, can, in principle, be detected in future runs, as shown on the right.}
\label{lowmasspbh}
\end{center}
\end{figure}

Note that, $M_{\rm PBH}$ here corresponds to the mass of a PBH at the formation epoch and disregards any further mass growth due to merging or accretion. Moreover, there exist various uncertainties associated with the numbers in this equation e.g. the efficiency factor $\gamma$ which is defined as the ratio between the mass collapsing into a PBH and the total mass associated to that mode within the horizon. Its value is usually taken as $\gamma \sim 0.2$ \cite{Carr:1975qj, Carr:2009jm} but there could be some uncertainties associated with the PBH collapse. Often, the effects of critical collapse are neglected wherein detailed numerical work has shown that the mass of PBHs formed after horizon reentry will depend on the amplitude of the overdensity $\delta$. Secondly a slightly smaller value of the radiation energy density today will lead to an ${\cal O}(1)$  difference in this relation. Finally, a slight difference arises due to the value of $g_\ast$, the relativistic number of degrees of freedom in the thermal bath when the mode responsible for the PBH re-enters the horizon, although the dependence of $M_{\rm PBH}$ on $g_\ast$ is weak. All these uncertainties could introduce a factor of ${\cal O}(10)$ in the final result so one should keep them in mind when comparing this relation with an exact numerical calculation, as is the case with our scenario.

In Figure \ref{lowmasspbh}, we have plotted the PBHs mass fraction at the formation epoch $\beta_{\rm ini}$ and the associated secondary GWs energy density for two different cases of our scenario \cite{Bhaumik:2019tvl}. 
Since the PBH mass fraction at matter-radiation equality $\beta_{\rm eq}$ can be expressed as $\beta_{\rm eq} \simeq (a_{\rm eq}/a_{\rm ini})\, \beta_{\rm ini}$, this implies that $\beta_{\rm eq}$ can be quite large even for a very small $\beta_{\rm ini}$. Since low mass PBHs will evaporate very quickly, $\beta_{\rm ini}$ is the only relevant quantity for such PBHs and thus, we have optimised our parameters\footnote{In order to obtain an appropriate fraction of initial PBH abundance, we need to fine-tune the potential parameters, following the same method, as in \cite{Bhaumik:2019tvl}, which requires slight deviations from the exact inflection point conditions of the inflationary potential.} such that we obtain $\beta_{\rm ini} \sim 10^{-23}$, as allowed by the constraints arising from the BBN and extragalactic $\gamma$-ray background. 
However, as pointed out recently in \cite{Ballesteros:2020qam} and we also observed it in our inflationary model, their ISGWB fall right in the design sensitivity curves of a future configuration of Advanced LIGO detector and thus, can be potentially detected. Moreover, this ISGWB also falls in the sensitivity contours of the proposed third generation ground based GWs detectors, Einstein Telescope (ET) and Cosmic Explorer (CE). Since this ISGWB overlaps with all these three future GWs detectors, there lies again an interesting possibility of its simultaneous detection with these GWs observatories and putting stringer constraints on its origin.  
Note that, the ground based GWs detectors such as LIGO and VIRGO  have already detected the astrophysical GWs signals from a few systems of binary black holes and neutron stars with large masses and will detect many more in near future. However, the characteristic shape of the GWs spectrum in these two cases is very different and thus, can be easily disentangled. 
Recently, some prospects of probing such ultralight PBHs using their ISGWB signatures with the Advanced LIGO detector have been discussed in \cite{Kapadia:2020pir}. 

%%%%%%%%%%%%%%%%%%%%%%%%%%%%%%%%%%%%%%%%%%%%%%%%%%%%%%%%

\section{A lower bound on the PBHs mass for USR inflation}
\label{low-mass-pbh}

In this section, we shall discuss how to obtain a stringent lower bound on the PBHs mass in the USR phase, both using an instantaneous transition as well as a smooth transition from the USR to the SR phase. We find that the lower bound obtained in both the scenarios are in agreement with each other. Moreover, our bounds also effectively apply even if there are brief deviations from SR after the USR phase. 
 
%%%%%%%%%%%%%%%%%%%%%%%%%%%%%%%%%%%%%%%%%%%%%%%%%%%%%%%%

\subsection{An instantaneous transition from USR to SR}
 
In order to estimate the lower bound, we first need the estimation of minimum number of e-folds in the final SR phase followed by a USR phase. Assuming that the inflaton is rolling in the positive direction, $\phi_{N} = d\phi/dN$ is positive. Since $\phi_{N}$ always decreases during the USR phase, let's assume that at $N=N_0$, $\phi_{N}$ reaches its minimum value. This can also be considered the point after which the SR potential takes over and $\phi_N$ starts to increase again. Note that, at this point, $\phi_{NN}=0$ and as the first SR parameter $\epsilon=\phi_{N}^2/ 2 M_{\rm Pl}^2 $ is very small, the second SR parameter $\eta = \epsilon - (\phi_{NN}/\phi_{N})$ is also very small and as a result, all the SR conditions are satisfied. Thus we can safely approximate the power spectrum of ${\cal R}_k$ with the SR result as
\begin{eqnarray}
P_{\cal R}(k) \simeq \frac{1}{8\pi^2 M_{\rm Pl}^2}\frac{H^2}{\epsilon_0},  \label{sl-pspec}
\end{eqnarray}
where $\epsilon_0 = \epsilon(N_0)$. Since we are considering an instantaneous transition from the USR to the SR phase, the smallest scale $k_0=k(N_0)$ leaving the horizon would still be amplified and produce PBHs. Moreover, a significant mass fraction of PBHs requires an amplification of the power spectrum at PBHs scales as $P_{\cal R}(k) \sim 10^{-2}$, so equation (\ref{sl-pspec}) leads to 
\begin{eqnarray}
\epsilon_0 \simeq \frac{H^2}{M_{\rm Pl}^2}.   \label{ep0}
\end{eqnarray}
Now, using the  SR approximation during the horizon exit of the observable pivot scale $k_p$, we can safely use equation (\ref{sl-pspec}) to estimate the Hubble parameter $H$ which stays nearly constant during inflation, as
\begin{eqnarray}
\frac{H^2}{M_{\rm Pl}^2} \sim  10^{-9} ,  \label{H0}
\end{eqnarray}
where we have used $P_{\cal R}(k_{p}) \sim  2.1 \times 10^{-9} $ and $\epsilon_{p} \sim  10^{-2}$ using $n_s(k_p) \simeq 0.965$. Now using (\ref{ep0}) and (\ref{H0}), we can estimate the minimum value of $\epsilon_0$ as $\epsilon_0 \sim 10^{-9}$ and the corresponding minimum value of $|{\phi_{N_0}}| \simeq  \sqrt{2} \times 10^{-4.5}$. Using this estimation, we want to understand the minimum number of e-folds necessary from $N=N_0$ to the end of inflation at $N=N_e$, where $\epsilon(N_e)=1$ or $\phi_{N_e} = \sqrt{2}$ must be satisfied. For this calculation, we shall assume that SR conditions are not violated again between the end of USR to the end of inflation and thus, the two conditions must be satisfied, $\epsilon \ll 1$ and $ |\eta| \le 1$. This leads to the following inequality  
\begin{eqnarray}
-1 \le \frac{\phi_{NN}}{\phi_N} \le 1 \quad {\rm or} \quad \l\vert  \frac{\phi_{NN}}{\phi_N} \r\vert \le 1. \label{con1}
\end{eqnarray}
If we assume  ${\phi_{NN}}/{\phi_N} =c(N)$, and solve it with the initial condition $\phi_N(N=N_0)={\phi_{N_0}}$, we obtain
\begin{eqnarray}
{\rm log}\l\vert\frac{\phi_N}{\phi_{N_0}}\r\vert= \int_{N_0}^{N_e}c(N) dN \label{sol1}
\end{eqnarray}
Now using (\ref{con1}), $|c(N)|\leq 1$,  so the minimum number of e-folds between $N_0$ to the end of inflation $N_e$ ($\phi_{N_e} = \sqrt{2}$) is constrained as;
\begin{eqnarray}
\Delta N = N_e -N_0 \geq {\rm log}\l\vert \frac{\sqrt{2}}{\phi_{N_0}}\r\vert .\label{sol2}
\end{eqnarray}
Using our previous estimation  $\l\vert{\phi_{N_0}}\r\vert \simeq \sqrt{2} \times 10^{-4.5}$, we get $\Delta N \geq 10.36$ which is roughly the duration of the final SR phase before the end of inflation.
 
%%%%%%%%%%%%%%%%%%%%%%%%%%%%%%%%%%%%%%%%%%%%%%%%%%%%%%%%

\subsection{A smooth transition from USR to SR}

Our previous estimation of the minimum number of e-folds was independent of the form of the potential and we only assumed an instantaneous transition from the USR to SR phase. To extend our analysis for a smooth transition from USR to SR, we need to consider the potential around $\phi(N_0)\equiv\phi_0$.
Since USR phase is on a flat part of potential, we can effectively approximate the potential around $\phi_0$, with first few terms of the Taylor's expansion as
\begin{equation}
V(\phi)={b_0}+{b_1} (\phi -{\phi_0})+{b_2} (\phi-{\phi_0})^2+ ....
\end{equation}
In the vicinity of the USR phase, we can neglect $\phi_{N}^2$ term and assume the Hubble parameter to be constant as $H(N) \simeq \sqrt{V(\phi_0)/3} = \sqrt{b_0/3}$. This reduces the equation of motion for $\phi$ to
\begin{equation}
\phi_{NN}+3 \phi_{N}+\frac{1}{H^2}\frac{d V}{d \phi}=0.
\end{equation}
Now, using the initial condition that at $N=N_0$ inflation field value is $\phi_0$ and the minima of $\phi_{N}$ is reached {\it i.e.}  $\phi_{NN}=0$, we obtain the dynamics of $\phi$ as \cite{Liu:2020oqe}
\begin{multline*}
\phi(N)=
\phi_0-\frac{{b_1}}{2
	{b_2}}+\frac{1}{4 \alpha {b2}}\times\\
\Bigg[ \left(\alpha  {b_1}+\frac{4 {b_1} {b_2}}{3 H}-3
	{b_1} H\right) e^{-\frac{(\alpha +3 H) (N-N_{0})}{2
			H}}+
		\left(\alpha  {b_1}-\frac{4 {b_1} {b_2}}{3 H}+3
	{b_1} H\right) e^{\frac{(\alpha -3 H) (N-N_{0})}{2
			H}} \Bigg],
\end{multline*}
where $\alpha= \sqrt{3 {b_0}-8 {b_2}}$. Assuming that the inflaton is rolling in the positive direction, at the minima, $\phi_{N}$ must be positive, to have a finite duration of the USR phase. So we need 
\begin{eqnarray}
\phi_{N}(N_0)=-\frac{{b_1}}{b_0}>0 \quad {\rm and} \quad \phi_{NNN}(N_0)=\frac{(6 b_1 b_2)}{b_0^2}>0.
\end{eqnarray}
These conditions constrain the possible value of potential parameters; $b_1<0$ and $b_2<0$. Using these, we can now express $\phi_{N}$ as a function of a single positive parameter $b_6$ as
\begin{equation}
\frac{\phi_{N}(N)}{\phi_{N}(N_0)}=\frac{\sqrt{3}}{b_6} e^{-3 {\Delta N}/2} \sinh \left(\frac{\sqrt{3}}{2} \,{b_6} {\Delta N}\right)
+e^{-3 {\Delta N}/2} \cosh \left(\frac{\sqrt{3}}{2} \, {b_6} {\Delta N}\right), \label{defy}
\end{equation}
where $b_6={\sqrt{3 {b_0}-8 {b_2}}}/{\sqrt{{b_0}}}$ and $\Delta N=N-N_0$.
To avoid eternal inflation, we need $b_2<0$, so the minimum value of $b_6$ must be greater than $\sqrt{3}.$ Now we want to estimate the number of  e-folds from the peak in the power spectra to the point where the SR potential takes over completely (let's say at $N=N_s$). Our assumption is that this transition to SR must happen while $|\eta| \le 1$ which then leads to 
\begin{align}
\epsilon(\Delta N)=&\,10^{-9}\,\l(\frac{\phi_{N}(\Delta N)}{\phi_N(N_0)}\r)^2,\\
\eta(\Delta N)=& \,\epsilon(\Delta N)-\frac{\phi_{NN}(\Delta N)}{\phi_N(\Delta N)} = \epsilon(\Delta N)+ \frac{9-3 {b_6}^2}{2 \sqrt{3}\, {b_6} \coth \left(\frac{\sqrt{3}}{2}  {b_6} {\Delta N}\right)+6}.  \label{eta1}
\end{align}
When the minima of $\phi_N$ is reached, $\phi_{NN}=0$ and $\epsilon$ is very small ($\epsilon \sim 10^{-9}$) so very close to $N=N_0$, $\eta$ crosses zero. Just after this crossing, $\phi_{NN}$ starts to increase and achieves a positive value which leads to an increase in $\phi_N$ as well. Very quickly, the ratio  $\phi_{NN}/\phi_N$ saturates, which is the second term of $\eta$ with a negative sign, as in (\ref{eta1}). It is interesting to note, that depending on the value of $b_6$, this term saturates to a negative asymptotic value $\eta_{\rm asym}=\frac{9-3 {b_6}^2}{2 \sqrt{3} {b_6}+6}$. Evidently, this saturation indicates the end of the transition phase. After this point, the further dynamics must be completely described by the SR potential.
If this saturation value $\eta_{\rm asym} \le -1 $, the dynamics deviates from SR before the transition and enters a fast roll phase. We shall assume that there is no intermediate fast roll phase between USR and SR and consider only values of $b_6$ which ensures the transition to SR before $\eta \simeq -1$. Thus solving for $\phi_{NN}/\phi_{N}=1$, we obtain
\begin{eqnarray}
N(b_6)=\frac{2}{\sqrt{3}\, {b_6}}  \coth ^{-1}\left(\frac{\sqrt{3} \left({b_6}^2-5\right)}{2 {b_6}}\right). \label{eq-N2}
\end{eqnarray}
In our case, $b_6$ is already constrained as $b_6>\sqrt{3}$, the above equation further limits the value of $b_6$ to be less than $5/\sqrt{3}$ so we finally have $ \sqrt{3} < b_6 \le 5/\sqrt{3}$.
 
Now we want to understand what can be the maximum value of $\epsilon_s$ or ${\phi}_{N_s}$ during this transition and how many e-folds are spent to reach the transition. If we solve $\eta=-c$ for $0<c<1$, from (\ref{eta1}) we get
\begin{equation}
N_c(b_6)=\frac{2}{\sqrt{3}\, {b_6}} \coth ^{-1}\left(\frac{\sqrt{3} \left({b_6}^2-2
		c-3\right)}{2 {b_6} c}\right). \label{N-c}
\end{equation} 
Upon using (\ref{N-c}), at $N=N_c$ we can find $\epsilon$ as a function of $b_6$ as
\begin{equation}
\epsilon(b_6)=3 \times10^{-9} \frac{\left({b_6}^2-3\right)}{3 {b_6}^2-(2 c+3)^2}\,  \exp \left[-\frac{2 \sqrt{3}}{b_6}
\coth^{-1}\left(\frac{\sqrt{3} \left({b_6}^2-2 c-3\right)}{2 {b_6} c}\right)\right]\label{ep-c}
\end{equation}
\begin{figure}[t]
\begin{center}
\includegraphics[width=7.9cm, height=5.8cm]{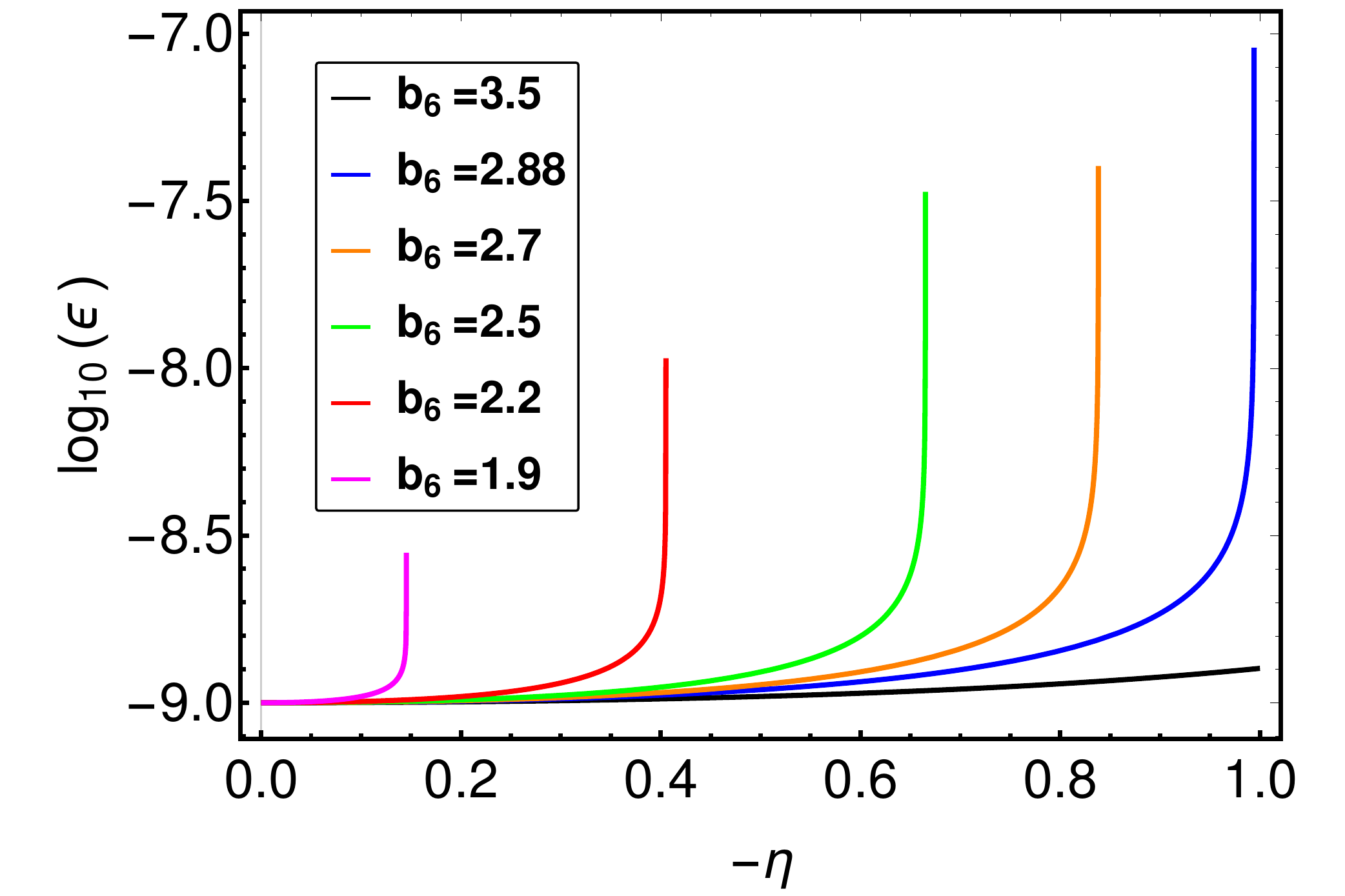}
\hspace{10pt}
\includegraphics[width=7.3cm, height=5.9cm]{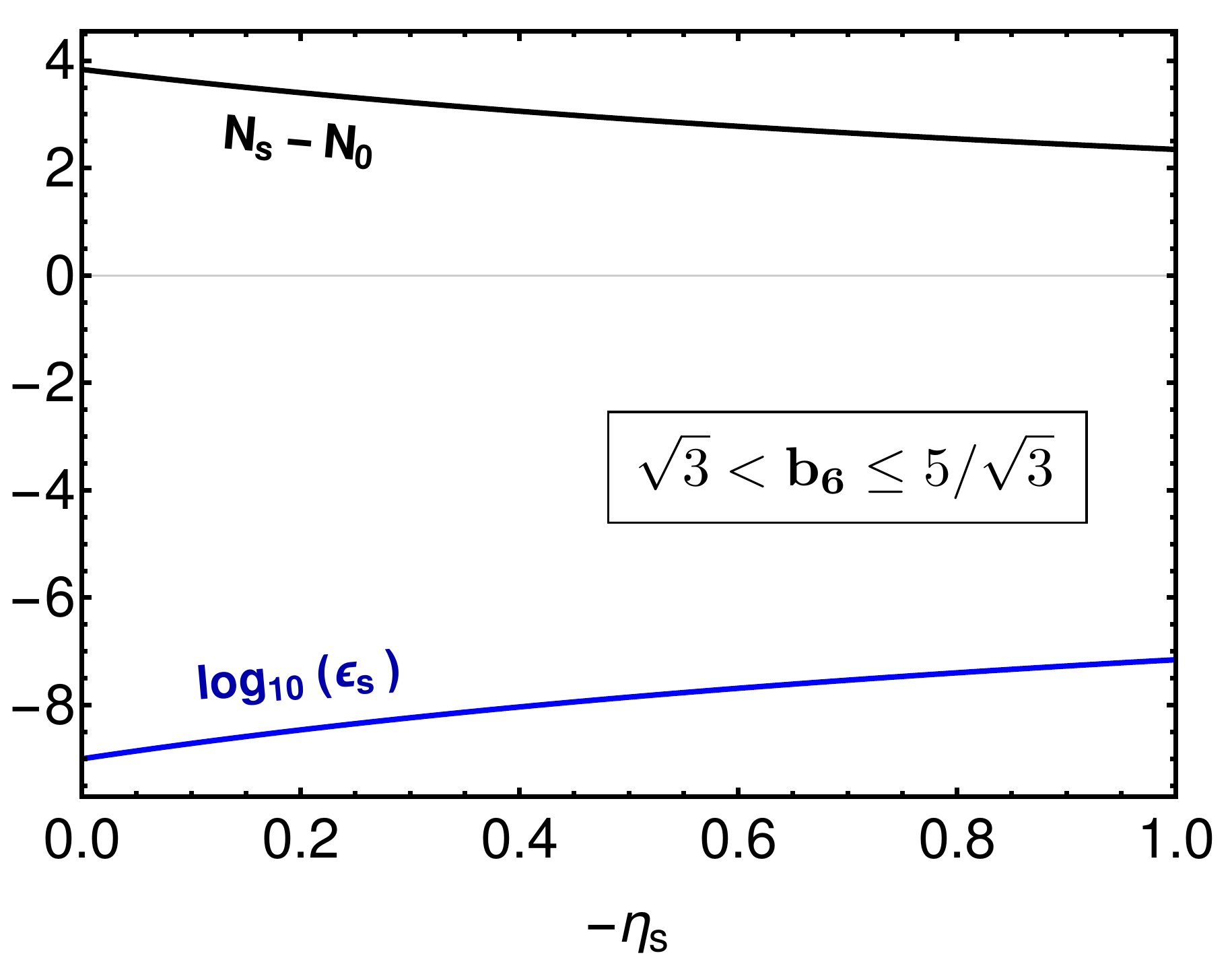}
\vskip 12pt
\caption{On the left, we have plotted the growing behaviour of $\epsilon$ before the transition, as obtained in equation (\ref{ep-c}). It is evident that for all values of $b_6$ in the range $ \sqrt{3} < b_6 \le 5/\sqrt{3}$, $\eta$ saturates to a constant value, greater than $-1$. On the right, the largest value of $\epsilon_s$ and the corresponding number of e-folds whenever $\eta$ reaches its asymptotic value $\eta_s$.}
\label{fig-low1}
\end{center}
\end{figure}
This value of $\epsilon$ is plotted in figure \ref{fig-low1} (on the left) and from this plot, we can clearly see that for every $b_6$ in the range $ \sqrt{3} < b_6 \le 5/\sqrt{3}$, $\eta$ saturates to a constant value, $\eta > -1$. We can consider this saturation as the beginning of the SR phase and we need to use the value of $\epsilon$ during this saturation as the starting value of $\epsilon_s$ for the SR phase. Solving for $\phi_{NN}/\phi_{N}=-\eta_{\rm asym}+\xi$ where $\xi$ is negative and much smaller than $\eta_{\rm asym}$, we can get the value of e-fold number where $\eta$ reaches its asymptotic value and the value of $\epsilon_s$, as plotted on the right in figure \ref{fig-low1}. It is evident from these plots that $ b_6 = 5/\sqrt{3}$ leads to the highest value of $\epsilon_s \simeq 7.0\times 10^{-8}$ and the lowest value of $N_{s}-N_0 \simeq 2.34721$. This also corresponds to the quickest transition and largest possible transition value of $|\eta_s|$ whereas  $ b_6 \to \sqrt{3}$ leads to the lowest value of $\epsilon_s \simeq 1.0\times 10^{-9}$ and largest value of $N_{s}-N_0 \simeq 3.83$.

Finally, the above values of $\epsilon_s$ can be translated to the bound on the number of e-folds using (\ref{sol2}) and thus, we can calculate the total number of e-folds, before and after the SR transition for both the cases as
\begin{enumerate}
\item 
For $ b_6 = 5/\sqrt{3}$, $\Delta N_{min} \simeq 2.35+8.23=10.58$
\item
For $ b_6 \to \sqrt{3} $, $\Delta N_{min} \simeq3.83+10.36=14.19$
\end{enumerate}
Thus, it is evident that for a  smooth transition, even quickest possible transition takes more number of e-folds than our estimation in the previous section wherein we considered an instantaneous transition in a rather model independent manner. 

%%%%%%%%%%%%%%%%%%%%%%%%%%%%%%%%%%%%%%%%%%%%%%%%%%%%%%%%

\subsection{Estimation of the lower bound on the PBHs mass}

In order to estimate a lower bound on the PBHs mass from the USR phase, we first need to calculate the total number of e-folds from the horizon exit of the pivot scale $k_p=0.05\, {\rm Mpc}^{-1}$ to the end of inflation. 
In a RD universe, $a\propto t^{1/2}$, $H \propto t^{-1}$ and $k=aH \propto t^{-1/2} \propto H^{1/2}$, thus if a comoving wavenumber $k$ re-enters the horizon during the RD epoch, we can relate it with the comoving mode $k_{\rm eq}$ of matter-radiation equality using the Hubble parameter during re-entry as  
\begin{align}
\frac{k}{k_{\rm eq}}=\left ( \frac{H}{H_{\rm eq}} \right)^{1/2}. \label{k-rat}
\end{align}
Upon using $\rho_{\rm eq} = 3 H_{\rm eq}^2 M_{\rm Pl}^2$, $H_{\rm eq}$ can be related to the present Hubble parameter as
\begin{align}
\rho_{\rm eq}=\,&2\, \Omega_{r,0} \rho_{c}(1+z_{\rm eq})^4  =6\, \Omega_{r,0} H_0^2 M_{\rm Pl}^2 (1+z_{\rm eq})^4 ,\\
H_{\rm eq} \simeq \,& \sqrt{\Omega_{r,0}} \,H_{0} (1+ z_{\rm eq})^2.
\end{align}
Taking $\Omega_{r,0} \simeq 8 \times 10^{-5}$, $z_{\rm eq} \simeq 3400$ and $H_{0} \simeq 2 \times 10^{-4}\, {\rm Mpc^{-1}}$, we find $H_{\rm eq} \simeq 20.7 \,{\rm Mpc^{-1}} $.
Note that, we are calculating $H_{\rm eq}$ by only taking into account the radiation energy density at the matter-radiation equality. Now at the pivot scale, we know both the amplitude and the tilt of the primordial spectrum from Planck so using (\ref{sl-pspec}), we can calculate $H_{\rm inf}$ during inflation which we assume to stay nearly constant up to the end of inflation which, for instantaneous reheating, is same as with the beginning of the RD epoch  so $H_{\rm inf} \simeq H_e \simeq H_{r}$ which is given by
\begin{eqnarray}
H_e \simeq H_{r} \simeq \left( 8\pi^2 M_{\rm Pl}^2 P_{\cal R}(k)\epsilon \right)^{1/2}.  \label{sl-Hp}
\end{eqnarray}
With this and (\ref{k-rat}), we can now calculate the comoving mode entering at the very beginning of the RD epoch, $k_{r}=a_{r}H_{r}$ as
\begin{eqnarray}
k_{r} \simeq \frac{k_{\rm eq}}{\sqrt{H_{\rm eq}}} \left( 8\pi^2 M_{\rm Pl}^2 P_{\cal R}(k)\epsilon \right)^{1/4}.  \label{k_ri}
\end{eqnarray}
Taking $k_{\rm eq} \simeq 0.01\, {\rm Mpc}^{-1}$ and $P_{\cal R}(k) \simeq 2.1 \times 10^{-9}$ at $k=k_p$, we find
\begin{eqnarray}
k_{r } \simeq 1.94\times 10^{24} \epsilon^{1/4}\, {\rm Mpc}^{-1}.
\end{eqnarray}
For the case of an instantaneous reheating, we can finally estimate the total number of e-folds from the horizon exit of the pivot scale to the end of inflation 
\begin{eqnarray}
N_e-N_p={\rm ln}\left( \frac{k_{r}}{k_p} \right) \simeq 58.92 + \frac{1}{4} {\rm ln}\, \epsilon. \label{N-expr-ep}
\end{eqnarray}
Using observational constraints on the value of $n_s \simeq 0.965$ at the pivot scale, we can roughly estimate $\epsilon \leq 10^{-2}$ which then limits $N_e-N_p \lesssim 57.77$. This is, of course, consistent with the results from \cite{Liddle:2003as} wherein the total number of e-folds from the horizon exit of the present horizon $(k_{\rm hor} \simeq 2 \times 10^{-4}\, {\rm Mpc}^{-1})$ to the end of inflation is constrained to be $\Delta N_{\rm tot} \simeq 63.3$.

%\subsection{Inference on the lower limit of PBH mass and upper limit of second order SGWB }
We now want to estimate the maximum number of e-folds from the horizon crossing of $k_p$ to the minima of $\phi_N$ at $N=N_0$ which we can translate to the smallest scale becoming super-horizon before the onset of the final SR phase. Assuming the Hubble parameter $H$ to be nearly constant and $k_p=a(N_p) H=a_ie^{N_p}H$ where $a_i$ is the scale factor at the beginning of inflation, using (\ref{sol2}) and (\ref{N-expr-ep}), we get
\begin{equation}
N_0-N_p \simeq 57.77-10.36=47.41 .\label{N-expr-ep-new}
\end{equation}
This corresponds to a scale  $k_{\rm PBH}=\sigma a(N_0) H$, which becomes super-horizon around $N=N_0$. Here $\sigma \ll 1$ is taken to ensure that super-horizon condition is satisfied 
\begin{equation}
k_{\rm max}=\sigma k_p e^{N_0-N_{p}} \approx 1.94 \times 10^{19} \sigma \, {\rm Mpc}^{-1}, \label{kmax}
\end{equation}
thus the smallest possible comoving length scale associated with a PBH peak corresponds to $k_{\rm max}$, or $k_{\rm PBH} \le k_{\rm max}$. Now we can use the relation between the mass of PBHs and the frequency of second order induced GWs as in equation (\ref{m-k-f-relation}) to translate the upper limit on $k$ in the above equation to a lower limit on the PBH mass, $M_{\rm PBH}$  and an upper limit on the frequency of corresponding second order GWs, $f_{}$ as
\begin{eqnarray}
M_{\rm PBH} &\ge& 6.14\times 10^{-27}\sigma^{-2} \,M_{\odot}\\
f &\le& 2.91 \times 10^4 \sigma\, {\rm Hz}
\end{eqnarray}
Now if we take a very conservative limit on the factor $\sigma$ as $\sigma \le 10^{-2}$, lowest possible value of PBH mass and the highest value of the second order GWs frequency turn out to be $M_{\rm PBH} \ge 6.14\times 10^{-23} M_{\odot} $ and $ f \le 2.91 \times 10^2$ Hz, respectively. Interestingly these constraints just cover the Advanced LIGO frequency range so future runs of Advanced LIGO can be used to detect the existence of PBHs from USR models, as we have already discussed in detail in Section \ref{aligo}.

Recently, it has also been discussed in the literature that during their formation, the abundance of PBHs with masses $ M < 10^9 \, {\rm g} \sim 5\times 10^{-25} M_{\odot} $ are essentially not constrained so they can constitute the dominant component of energy density (since they behave as matter) and drive the dynamics of the universe for a brief period of time before their evaporation due to Hawking radiation \cite{Inomata:2020lmk}. However our analysis and results of these sections suggest that USR models of inflation can not possibly produce such ultra low mass PBHs to dominate the energy density of universe for even a short period of time. Thus one has to resort to a different mechanism of PBHs production to discuss those possibilities. It may be interesting to see if an intermediate fast roll phase can produce such ultra low mass PBHs which can dominate for a short while before their evaporation. 

%%%%%%%%%%%%%%%%%%%%%%%%%%%%%%%%%%%%%%%%%%%%%%%%%%%%%%%%

\section{Imprints of reheating on lower mass bound and ISGWB spectrum}
\label{reh-gw}

In the previous section, we had derived a lower mass bound of PBHs originating from a class of USR models of inflation, and also estimated the induced GW energy density coming from the inflection point model we have studied in our previous paper. Both these results have strong dependence on the reheating history between the end of inflation and the start of the RD phase. Our discussion, so far, was limited to an instantaneous reheating. In this section, we shall extend our results by taking into account the effects of a non-instantaneous reheating stage. 

\subsection{Effects of reheating on the lower mass bound}
In order to understand the effects of reheating on the PBHs mass bound, we shall first parameterize the reheating phase with an equation of state parameter $w_{\rm reh}$ and its duration in number of efolds $N_{\rm reh}$.  We also assume a sudden transition from the reheating phase to the RD epoch. Following the same arguments as in the previous section, a non-instantaneous reheating epoch shall modify equation (\ref{N-expr-ep-new}), to 
\begin{equation}
N_0-N_p \simeq 47.41 -\frac{1}{4}N_{\rm reh}(1-3w_{\rm reh}),  \label{N-expr-ep-reh}
\end{equation}
which implies that 
\begin{equation}
k_{\rm max} \simeq  1.94 \times 10^{19}\, \sigma\, e^{-\frac{1}{4}N_{\rm reh}(1-3w_{\rm reh})}\, {\rm Mpc}^{-1}.
\label{kmaxx}
\end{equation}
Note that, both $N_{\rm reh}=0$ or $w_{\rm reh}=1/3$ correspond to instantaneous reheating. 
Since the maximum value of $k$ shifts to an even lower value, both the lower bound on the PBHs mass and the upper bound on the GWs frequency become stronger and are given by
\begin{eqnarray}
M_{\rm PBH} &\ge& 6.14\times 10^{-27}\sigma^{-2} e^{\frac{1}{2}N_{\rm reh}(1-3w_{\rm reh})}\, M_{\odot}, \\
f &\le& 2.91 \times 10^4 \sigma\, e^{-\frac{1}{4}N_{\rm reh}(1-3w_{\rm reh})}\, {\rm Hz}.
\end{eqnarray}
For $\sigma \le 10^{-2}$, $w_{\rm reh}=0$ and $N_{\rm reh}=10$, these bounds translate to $M_{\rm PBH} \ge 9.09\times 10^{-21} M_{\odot} $ and $f \le 23.94\, {\rm Hz}$.
So evidently the lower bounds calculated in the previous section become even stronger in the presence of a non-zero duration of the reheating phase. 

\subsection{Effects of reheating on the ISGWB}
\label{rehgw}
To consider the broad effects of different reheating histories on the ISGWB originating from our inflection point inflationary model, we shall  assume an effectively sudden transition from the reheating phase to the RD epoch. With this setup, we notice that a non-instantaneous reheating history leaves two very different effects on the observable energy density of ISGWB.

We have discussed the origin of the first effect in details in our previous paper \cite{Bhaumik:2019tvl}.  For any inflationary model, different reheating histories change the mapping of different length scales upon their re-entry to the horizon which shifts the peak of the scalar power spectra to lower wavenumbers $k$ (figure 4 of \cite{Bhaumik:2019tvl}). It also slightly changes the pivot scale normalisation, thereby leading to more abundant PBH formation in higher mass range. As the ISGWB involves a convolution integral of first order scalar power spectra, this shift in the peak of the scalar spectra is reflected as a shift towards lower frequencies in the ISGWB energy density. We found that this effect is strongest for a matter dominated reheating ($w_{\rm reh}=0$) so if we consider the expression of $\Omega_{GW}$ as in (\ref{omega-in-u-v}), this effect is completely encoded through the primordial scalar power spectra.

The second effect arises due to the nontrivial evolution of scalar perturbation modes and is only significant for an eMD reheating phase. While an eMD epoch leads to a constant transfer function for first order scalar perturbations, for all subhorizon modes, any other reheating phase (with $w_{\rm reh}>0$) would lead to a suppression of the transfer function far before the RD starts. A similar suppression for all the subhorizon modes will happen if the transition from the eMD to RD phase is slow. While we shall not observe any  amplification of $\Omega_{GW}h^2$ for $w_{\rm reh}>0$ cases and gradual transition from eMD to RD, a sudden transition from eMD to RD will actually lead to a secondary amplification. For the case of a nearly scale invariant scalar power spectra, this effect is discussed in detail in \cite{Inomata:2019ivs}.

In an eMD reheating phase, the first order scalar perturbations for both the sub and super horizon scales, stay nearly constant. Just after the transition to RD, the amplitude of all these modes oscillates rapidly and they quickly decay. The oscillation frequency for each mode depends on the corresponding wavenumbers. Both the transfer function $\mathcal{T}(x,x_r)$ and its time derivative contribute to the term  $f(u,v,x,x_r)$ as we can see from  (\ref{ff}). So for modes with very large wavenumbers which re-enter the horizon during the eMD phase, the terms involving the time derivative of $\mathcal{T}(x,x_r)$ in (\ref{ff}) contribute dominantly, leading to a secondary amplification of the ISGWB spectra and the the frequency of maximum amplification corresponds to the cutoff scale $k_{\rm max}$ of the scalar power spectra. In our case, we take $k_{\rm max}$ to be the end of inflation scale as any scale smaller than this always remains sub-horizon and thus, no growth can happen for those scales. This effect is essentially encoded in the integral $I(u,v,x,x_r)$, as this term covers the time evolution of the source function. We can split the contribution of $I$ integral into two parts as
\begin{equation}
I=I_{eMD}+I_{RD},
\end{equation}
where $I_{eMD}$ and $I_{RD}$ are the contributions to the ISGWB produced during eMD and during RD, respectively. The calculation of $I_{eMD}$ strongly depends on the gauge choice and recently it has been argued that, during a phase of $w \leq 0$, one can neglect this contribution entirely by taking a suitable gauge choice \cite{Domenech:2020xin}. Also, for the conformal Newtonian gauge, the magnitude arising from $I_{eMD}$ is sufficiently lower than the contribution of $I_{RD}$ \cite{Inomata:2019ivs} so in our calculations, we neglect any contribution, coming from the $I_{eMD}$ part and focus solely on the contribution from $I_{RD}$ to the ISGWB produced after the transition to RD.

\begin{figure}
\begin{center}	
\includegraphics[scale=0.33]{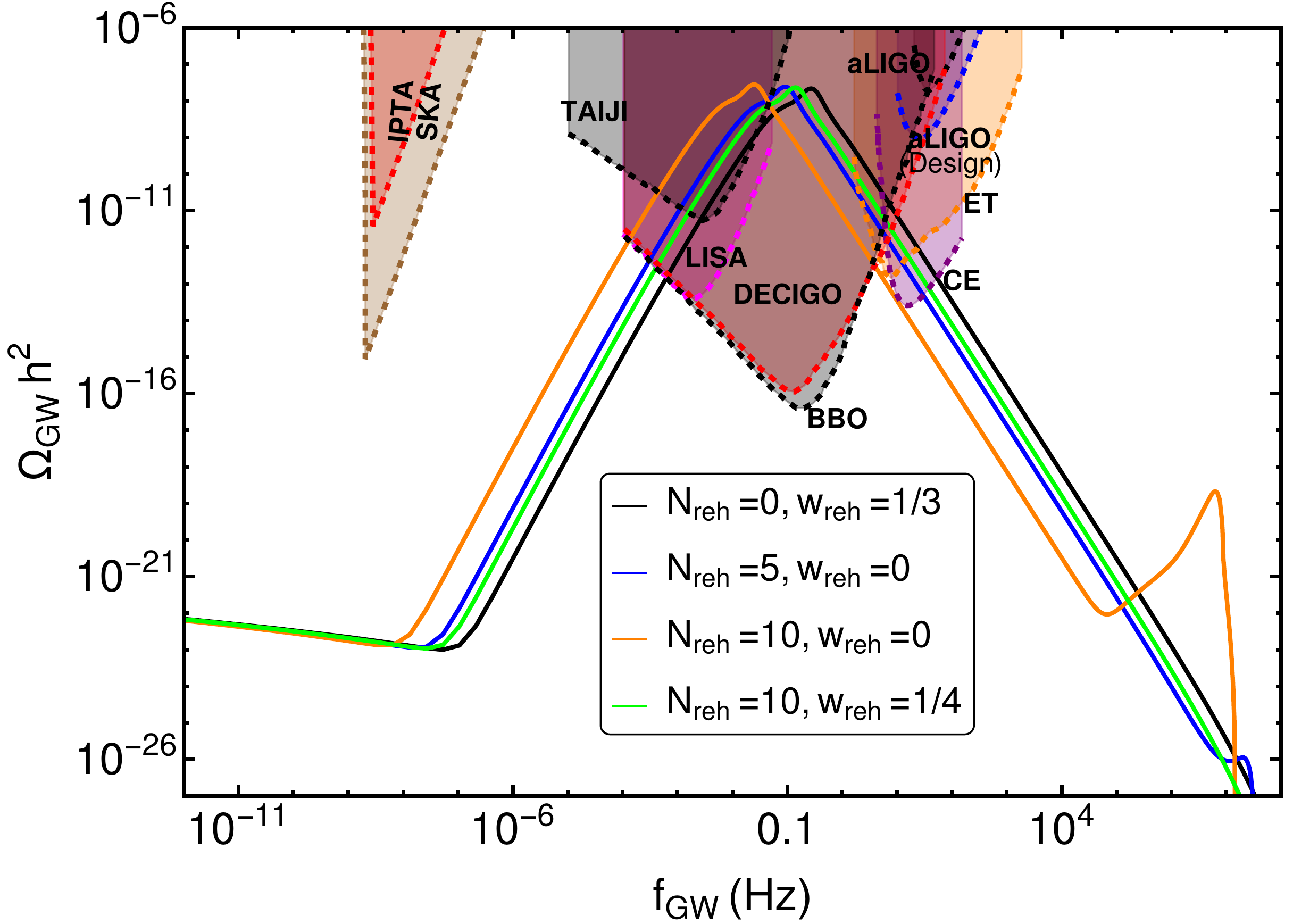}
\hspace{10pt}
\includegraphics[scale=0.28]{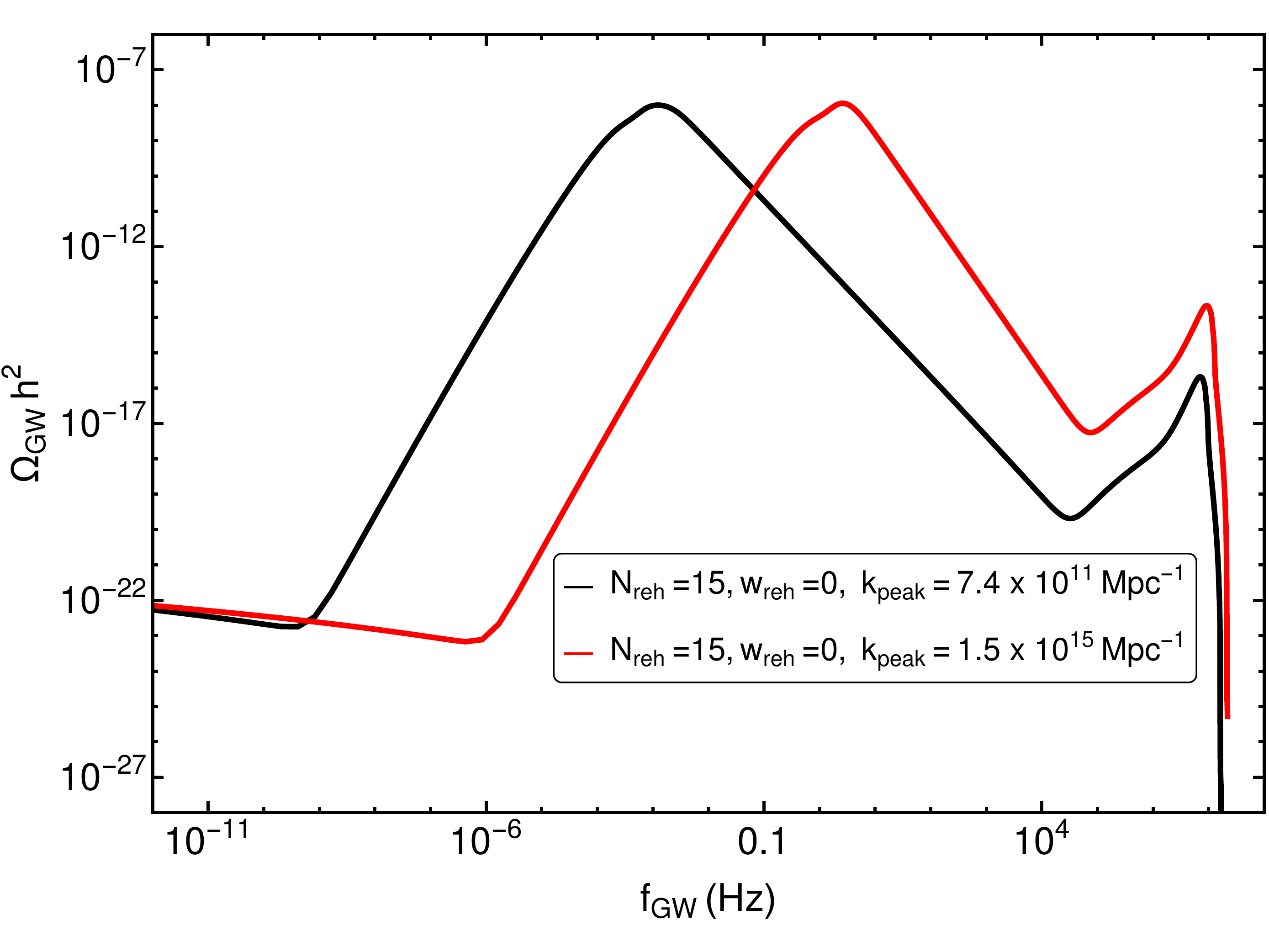}
\vskip 12pt
\caption{On the left, the ISGWB energy density $\Omega_{GW}h^2$ has been plotted for different reheating histories, as shown in the inset. On the right, the ISGWB energy density for same reheating history but for two different scalar power spectra peaking at different wavenumbers $k$, leading to same amplitude at different frequencies for the first peak but different amplitudes at the same frequency for the second peak.}
\label{reh1}
\end{center}
\end{figure}

For a RD phase, preceded by an eMD era, an analytical expression of $I_{RD}(u,v,x,x_r)$ integral can be obtained after integrating each term by parts. The full analytical expression of $I(u,v,x,x_r)$, (given in appendix \ref{appendix}) is used to get the ISGWB energy density numerically. We also match the numerical results with the corresponding results in limiting cases, $x_r \ll 1$ and $x_r \gg 1$. These two different limits represent the  two different peaks quite closely i.e. $x_r \ll1$ (or $k \ll k_r$) limit corresponds to the pure RD era result, and leads to the first peak of ISGWB in figure \ref{reh1} while  $x_r \gg 1$ (or $k \gg k_r$) limit correctly represents the second peak. The expressions involving the second limit are derived in (\ref{ciclxr}) and (\ref{cislxr}) of appendix \ref{appendix} while the expressions for the first limit would reduce to the standard expression (\ref{ird}) for pure RD case. 

As seen in the left panel of figure \ref{reh1}, any reheating history, different than instantaneous one, leads to a shift in the primary peak of the ISGWB power spectra which comes due to the first effect. The second peak only occurs for an eMD reheating and frequency of this peak depends on the reheating history. 
To demonstrate this point, we have also considered the case of $w_{\rm reh}=1/4$ which is an intermediate state of fluid between the radiation ($w_{\rm reh}=1/3$) and the matter ($w_{\rm reh}=0$).  We found that the second bump does not appear in the case of $w_{\rm reh}=1/4$ as the behaviour of the transfer function in this case is very similar to $w_{\rm reh}=1/3$.
Note that, the frequency of this second resonant peak corresponds to the cutoff scale of the power spectra $k_{\rm max}$, which we can take as the smallest comoving scale leaving the horizon during inflation, or the end of inflation scale. As we have discussed in the beginning of this section, different reheating histories shall lead to different values of $k_{\rm max}$ [c.f. equation (\ref{kmaxx})] and thus slightly different values of second peak frequency. It is this dependence, which is reflected here, in the slight shift in frequency for the second resonant peak in two different durations of the eMD phase. The understanding of the amplitude of the second resonant peak is slightly more complicated, as it depends both on the duration of the reheating phase and also on the amplitude of the primordial scalar power spectra around the cutoff scale and this amplitude strongly varies as the frequency of the primary peak changes. As we see in the right panel of figure \ref{reh1}, for the same duration of eMD epoch, if we take the power spectra with peaks at two different wavenumbers (producing PBHs in two different mass range and primary peak in GW spectra at two different frequencies), the amplifications of the second resonant peak are very different.

%%%%%%%%%%%%%%%%%%%%%%%%%%%%%%%%%%%%%%%%%%%%%%%%%%%%%%%%

\section{Conclusions and discussions  }
\label{discuss}

In this paper, we have studied the ISGWB from the enhancement of primordial curvature perturbations at smaller scales. An enhancement of the power spectrum is a very generic feature of all the inflationary models allowing the violation of slow roll conditions \cite{Jain:2007au, Jain:2008dw, Jain:2009pm}.  In our previous work \cite{Bhaumik:2019tvl}, we had presented an inflationary scenario with a polynomial potential containing an inflection point which can generate PBHs in different mass windows with a nearly monochromatic mass fraction. In particular, PBHs generated in the asteroid mass window are very interesting as first, they can contribute to the entire CDM in the universe and second, the induced GWs have a characteristic peak around the mHz frequency band which can be probed by the future space based GWs observatories such as LISA, DECIGO or BBO. We have also shown that the secondary GWs induced by more massive PBHs which will be peaked in the lower frequency range can be probed by IPTA/SKA observations. 
Interestingly, in this scenario, we also notice that very light PBHs which may completely evaporate by today and would not contribute to the dark matter at all, will also generate an ISGWB that may be observed from a future design of the ground based Advanced LIGO detector.

Further, using a model independent approach, we have obtained a lower bound on the PBHs mass by only assuming an instantaneous and a smooth transition from the USR to the SR phase. The lower mass bound of PBHs also become stronger in the case of a non-instantaneous reheating epoch. 
We also investigate the effects of reheating on the ISGWB spectrum and find that an epoch of a non-instantaneous reheating can cause a shift in the GWs spectrum to smaller frequencies, thereby making it accessible to the reach of different GWs observatories. In particular, we found that a transition from an eMD phase to a RD phase leads to a secondary enhancement of the ISGWB energy density spectrum on much larger frequencies. 
We stress that our motivation behind the calculation of the lower mass bound is two-fold. First, this mass bound is a model independent bound and thus applicable for a large class of USR models, similar to ours. Second, this result has very interesting correlations with our discussion of the very low mass PBH in Sec. \ref{aligo} and its relevance in determining the scales corresponding to the second peak in the induced GW background in the matter dominated reheating epoch in Sec. \ref{rehgw}. The first step of the lower mass bound calculation is to determine the smallest scale that leaves the horizon just before the end of inflation. It offers us an intuitive way to understand the location of the second GW peak for a matter dominated reheating. The second step in this calculation is to get a bound on the smallest possible scale leaving the horizon just before the onset of the final slow roll phase such that it leads to a PBH peak in primordial scalar power spectra. This provides us a lower mass limit on the ultra light PBHs. It also limits how close the first GW peak (corresponding to the PBH scale) can be to the second GW peak (end of inflation scale). As shown in the right panel of figure \ref{reh1}, the amplitude of the second GW peak strongly depends on the proximity to the first GW peak so this, in principle, also limits the amplitude of the second GW peak.

In general, all cosmological sources of GWs typically produce stochastic backgrounds of GWs with a frequency roughly related to the size of the comoving Hubble horizon at the time of their production. It is worth pointing out that the entire mechanism of PBHs generation from scalar field inflationary models\footnote{It is well known that dynamical gauge fields during inflation provide very rich and interesting phenomenology \cite{Durrer:2010mq, Byrnes:2011aa, Jain:2012ga, Jain:2012vm, Maleknejad:2012fw, Ferreira:2013sqa, Ferreira:2014hma}. In models wherein PBHs are produced due to the amplification of gauge fields, there exists another primordial contribution to tensor perturbations sourced directly by enhanced gauge fields during inflation \cite{Ozsoy:2020ccy, Ozsoy:2020kat}.} leads to different characteristic backgrounds of GWs which can be distinguished based on their spectral energy density and frequency range.
\begin{itemize}
\item 
Primordial GWs background from vacuum tensor fluctuations characterised by the tensor to scalar ratio $r$. This background is typically very small and highly redshifted since its generation during inflation thus can not be directly detected with present or future GWs observatories
\item
Secondary GWs background sourced by the enhanced scalar perturbations in models of PBHs formation. Such GWs production is
maximized when the scalar modes re-enter the horizon during the RD era but decay inside the horizon. This induced contribution typically has a broad peak in the spectra energy density which can be probed with various ground and space based GWs detectors.
\item
The GWs produced by the mergers of PBHs binaries since formation until today \cite{Clesse:2016ajp, Mandic:2016lcn}. The  frequency of this GWs signal is in the Hz - kHz regime which falls in the sensitivity band of ground based detectors such as LIGO and future runs such as O5 of Advanced LIGO. Perhaps, the binary black hole systems detected by LIGO are PBHs binaries. The most recently detected merger event \cite{Abbott:2020tfl} also points out to a strong possibility of these intermediate mass black hole's origin being primordial \cite{Abbott:2020mjq}.
\item
The GWs are also produced due to the graviton emission from Hawking evaporation of PBHs \cite{Inomata:2020lmk}. The emitted GWs from tiny PBHs with high Hawking temperature typically have very high frequency and thus, quite far from the reach of near future GW observatories. 
\end{itemize}

Recently, it has been discussed whether the spectral energy density of ISGWB in the RD era from first order scalar perturbations is gauge invariant. In principle, a physical observable today should not depend on the choice of the gauge in which the calculations are carried out. There have been a few papers discussing this issue lately \cite{Tomikawa:2019tvi, DeLuca:2019ufz, Inomata:2019yww, Yuan:2019fwv, Lu:2020diy} and all of them seem to present different conclusions. All these papers have computed the spectral energy density of induced gravitational waves in the Newtonian, comoving and the uniform curvature gauges. In \cite{Tomikawa:2019tvi}, it was noticed that there are huge differences in the final result between the Newtonian and the comoving gauge while the uniform curvature gauge gives the same result as the Newtonian gauge. However, in \cite{DeLuca:2019ufz, Inomata:2019yww, Yuan:2019fwv}, it was claimed that the induced GWs today are gauge invariant while ref. \cite{Lu:2020diy} claims that the result is identical in four different gauges but still different than other gauges. In summary, the issue of gauge invariance for second order GWs is not yet completely settled and requires further investigation. 

Having computed the power spectrum of induced tensor perturbations,  it is interesting and relevant to understand the extent of the induced tensor bispectrum in such models  and analyse whether its imprints could possibly be detected by future space based GWs observatories. Recently, it has been pointed out that the non-Gaussianity associated with the induced tensor bispectrum in some models can be large \cite{Bartolo:2018evs, Bartolo:2018rku} and it is imperative to think about the extent of this bispectrum in other inflationary models. Moreover, one can naively expect that all such models which induce a large ISGWB due to the enhancement of primordial curvature perturbations will also generally induce a large tensor bispectrum at the time of GWs production i.e. some time after the horizon re-entry of different modes.  However, it has been further emphasised that this peculiar non-Gaussian characteristic of the signal may not be observable unfortunately in any GWs detector at present. Since the detectors can only measure the superposition of such signals coming from many different directions in the sky and not just from one line of sight, such non-Gaussianities (or the phase correlations) would be further de-correlated by the propagation of GWs from different directions due to the inhomogeneities present from their generation epoch to today and thus, will not be observable \cite{Bartolo:2018evs}. 

It has been pointed out recently that anisotropies in the GWs background are an interesting observable that can be used to distinguish among different GWs production mechanisms \cite{Bethke:2013aba, Geller:2018mwu}. These anisotropies refer to a change in the spectral energy density of observed GWs as a function of direction in the sky and can be imprinted both at the generation epoch as well as due to their propagation through the perturbed universe from the formation epoch to today. These anisotropies are similar to the CMB anisotropies and can be computed using a Boltzmann approach taking into account both the scalar and tensor perturbations \cite{Contaldi:2016koz, Bartolo:2019oiq, Bartolo:2019yeu}. Recently, the effects of primordial curvature perturbations on  GWs propagation over cosmic distances have been calculated and it was shown that the resulting deformations of the GWs background can be significant for extremely peaked GWs spectra \cite{Domcke:2020xmn}. It will be very interesting to study these anisotropies in the case of scalar induced GWs background and see if they provide further insights into the mechanism of PBHs formation and the associated secondary GWs background produced in the early universe \cite{Bartolo:2019zvb}. 

Detecting very high frequency GWs is going to be a big challenge for future detectors as high frequency poses severe complications for interferometric observatories. An interesting thought in this context is based on an indirect detection of these high frequency GWs by means of their conversion into electromagnetic (EM) radiation (gravitons $\to$ photons) in the presence of a cosmological background magnetic field. This effect is often called the inverse Gertsenshtein effect \cite{Gertsenshtein:1962lgw, Zel'dovich:1974JETP}. It has been discussed that relic gravitons emitted by PBHs prior to BBN would transform to an almost isotropic background of electromagnetic radiation due to their conversion \cite{Dolgov:2011cq, Ejlli:2019bqj, Domcke:2020yzq}. This can be calculated at the recombination epoch and during the subsequent evolution of the universe. Since the produced EM radiation is concentrated in the X-ray part of the spectrum, this contribution could be observable and even dominate the cosmic X-ray background.
We plan to investigate all such interesting issues in future.

%%%%%%%%%%%%%%%%%%%%%%%%%%%%%%%%%%%%%%%%%%%%%%%%%%%%%%%%

\section*{Acknowledgment}
We would like to thank Chris Byrnes and Ranjan Laha for useful discussions. NB thanks Jishnu Sai P and Sahel Dey for fruitful discussions.  The financial support from the new faculty seed start-up grant of IISc, the Core Research Grant CRG/2018/002200 from the Science and Engineering Research Board, Department of Science and Technology, Government of India and the Infosys Foundation, Bangalore is greatly acknowledged. 

%%%%%%%%%%%%%%%%%%%%%%%%%%%%%%%%%%%%%%%%%%%%%%%%%%%%%%%%

\section{Appendix A : Calculation of $I_{RD}(u,v,x,x_r)$ for a RD epoch preceded by an eMD epoch}
\label{appendix}
 The general solution for the transfer function in RD, preceded by an eMD phase can be written as \cite{Inomata:2019ivs}
\begin{equation}{\label{phik}}
\mathcal{T}(x,x_r)=\frac{3 \sqrt{3}  \left[A(x_r)\, j_1 \left(\frac{x-x_r/2}{\sqrt{3}} \right)+B(x_r)\, y_1 \left(\frac{x-x_r/2}{\sqrt{3}} \right) \right]}{x-x_r/2},
\end{equation}
where $j_1$ and $y_1$ are the spherical Bessel functions of first and second kind, $x_r= k \tau_r$ and $x=k \tau $, where $\tau_r$ corresponds to the conformal time at the transition from eMD to RD, and $\tau$ corresponds to some conformal time after the transition, and $A, B$ are constants which depend on the duration of the eMD phase. Demanding the continuity of the transfer function $\mathcal{T}(x)$ and its time derivative at the transition, we can determine $A$ and $B$ as
\begin{equation}
\begin{aligned}{\label{aa}}
A(x_r)& = \frac{x_r}{2\sqrt{3}}\sin  \left(\frac{x_r}{2\sqrt{3}} \right)-\frac{1}{36}  (x_r^2-36 ) \cos  \left(\frac{x_r}{2\sqrt{3}} \right),\\
%\begin{equation}{\label{bb}}
B(x_r)& =-\frac{1}{36}  (x_r^2-36 ) \sin  \left(\frac{x_r}{2\sqrt{3}} \right)-\frac{x_r}{2\sqrt{3}} \cos  \left(\frac{x_r}{2\sqrt{3}} \right),
\end{aligned}
\end{equation}
where the pure RD case can be recovered by taking $x_r \to 0$ limit. For this general expression of transfer function, we can calculate $f(u,v,x)$
%\begin{equation}\label{ff}
\begin{align}\label{ff}
f(u,v,\bar{x},x_r)=\frac{4}{9} \biggl[ (\bar{x}-x_r/2) \partial_{\bar{x}}\mathcal{T}(u \bar{x},u x_r) ( (\bar{x}-x_r/2)\partial_{\bar{x}}\mathcal{T}(v \bar{x},vx_r) +\mathcal{T}(v \bar{x},v x_r)) \biggr. \\
+ \biggl. \mathcal{T}(u \bar{x},u x_r) ((\bar{x}-x_r/2)
\partial_{\bar{x}}\mathcal{T}(v \bar{x},v x_r)+3 \mathcal{T}(v \bar{x},v x_r)) \biggr].
\end{align}
%\end{equation}
The propagation of GWs shall not be affected by any preceding phase of eMD, so the Green's function for (\ref{EOM_Green}) shall be same as pure RD, and we can define the integral $I(u,v,x,x_r)$, running from the start of RD $x_r$ to a later time in RD by which the source term becomes inactive, as
\begin{equation}{\label{II}}
\begin{aligned}
I(u,v,x,x_r)= & \int_{x_r}^x d\bar{x} \frac{a(\bar{x})}{a(x)}f(u,v,\bar{x},x_r) k G(\bar{x},x),
   \end{aligned}
\end{equation}
where
\begin{align}
\frac{a(\bar{x})}{a(x)}=&\frac{\bar{x}-x_r/2}{x-x_r/2}
\end{align}
For simplifying the calculations, we can take $1/(x-x_r/2)$ out of the integral expression and define
\begin{equation}{\label{CI}}
\mathcal{I}(u,v,x,x_r)=I(u,v,x,x_r)\times (x-x_r/2) 
\end{equation}
We can now expand the integrand according to the powers of $\bar{z}=\bar{x}-x_r/2$ and integrate by parts seperately to obtain the analytical expression of  $\mathcal{I}(u,v,x,x_r)$ which can be broken into 6 different terms as
\begin{equation}{\label{brk-CI}}
\begin{aligned}
 \mathcal{I}(u,v,x,x_r) & =  \mathcal{I}_s(u,v,x,x_r) \sin(x) + \mathcal{I}_c(u,v,x,x_r) \cos(x) \\
 & + \mathcal{I}_{sm}(u,v,x,x_r) \sin  \left(\frac{x (u-v)}{\sqrt{3}} \right) + \mathcal{I}_{sp}(u,v,x,x_r) \sin  \left(\frac{x (u+v)}{\sqrt{3}} \right) \\
 & + \mathcal{I}_{cm}(u,v,x,x_r) \cos  \left(\frac{x (u-v)}{\sqrt{3}} \right) + \mathcal{I}_{cp}(u,v,x,x_r) \cos  \left(\frac{x (u+v)}{\sqrt{3}} \right).
 \end{aligned}
\end{equation}
 To calculate the ISGWB energy density spectra, we need to take the oscillation average of the square of $\mathcal{I}(u,v,x,x_r)$. In the late time limit, $x \gg 1 $, we can neglect all the terms except for the first two terms which simplifies the calculation to
\begin{equation}{\label{CIsqr}}
\begin{aligned}
 \overline{\mathcal{I}^2}(u,v,x,x_r) =\frac{1}{2} \left( \mathcal{I}_{s}^2(u,v,x,x_r) + \mathcal{I}_{c}^2(u,v,x,x_r)\right).
 \end{aligned}
\end{equation}
Now, we can further break $\mathcal{I}_s(u,v,x,x_r)$ and $\mathcal{I}_c(u,v,x,x_r)$ terms into two parts, one involving the $\text{Si}()$ and $\text{Ci}()$ integrals and another without them. 
\begin{equation}{\label{cics}}
\begin{aligned}
 \mathcal{I}_s(u,v,x,x_r) &=  \mathcal{I}_{s1}(u,v,x,x_r) + \mathcal{I}_{s2}(u,v,x,x_r) \\
\mathcal{I}_c(u,v,x,x_r) & = \mathcal{I}_{c1}(u,v,x,x_r) + \mathcal{I}_{c2}(u,v,x,x_r) \\
 \end{aligned}
\end{equation}
In $ (x - x_r/2) \gg 1$ limit, using $\lim_{x \to \pm \infty} \text{Ci}(x-x_r/2) \to 0$ and  $\lim_{x \to \pm \infty}\text{Si}( x-x_r/2) \to \pm \pi/2$, we can obtain simplified expression for these 4 terms. In this limit, we define $\text{Si}\left( (-1+ \frac{(u+v)}{\sqrt{3}})(x-x_r/2)\right) =  \mathcal{V} $ so that, for $\left(-1+ \frac{(u+v)}{\sqrt{3}}\right) > 0 $; $ \mathcal{V}=\pi/2$ and for $\left(-1+ \frac{(u+v)}{\sqrt{3}}\right) < 0 $; $ \mathcal{V}=-\pi/2$. These different terms in their simplest form can be expressed as\\

%\begin{equation}{\label{cis1}}\\
$\mathcal{I}_{s1} =$
%\breqnsetup{breakdepth={1}}
\begin{dmath}{\label{cis1}}
\frac{3}{8u^3 v^3} \left(-3+u^2+v^2\right)^2 \left(-\left((\pi +2 \mathcal{V}) A(v x_r) \left(B(u x_r) \text{Cos}\left(\frac{x_r}{2}\right)+A(u
x_r) \text{Sin}\left(\frac{x_r}{2}\right)\right)\right)-B(v x_r) \left((\pi +2 \mathcal{V}) A(u x_r) \text{Cos}\left(\frac{x_r}{2}\right)+(3
\pi -2 \mathcal{V}) B(u x_r) \text{Sin}\left(\frac{x_r}{2}\right)\right)+2 \text{Ci}\left(\frac{1}{6} \left(-3+\sqrt{3}
u-\sqrt{3} v\right) x_r\right) \left(-\left((A(u x_r) A(v x_r)+B(u x_r) B(v x_r)) \text{Cos}\left(\frac{x_r}{2}\right)\right)+(A(v
x_r) B(u x_r)-A(u x_r) B(v x_r)) \text{Sin}\left(\frac{x_r}{2}\right)\right)+2 \text{Ci}\left(-\frac{1}{6}
\left(3+\sqrt{3} u-\sqrt{3} v\right) x_r\right) \left(-\left((A(u x_r) A(v x_r)+B(u x_r) B(v
x_r)) \text{Cos}\left(\frac{x_r}{2}\right)\right)+(-A(v x_r) B(u x_r)+A(u x_r) B(v
x_r)) \text{Sin}\left(\frac{x_r}{2}\right)\right)+2 \text{Ci}\left(\frac{1}{6} \left(-3+\sqrt{3} u+\sqrt{3} v\right) x_r\right)
\left((A(u x_r) A(v x_r)-B(u x_r) B(v x_r)) \text{Cos}\left(\frac{x_r}{2}\right)-(A(v
x_r) B(u x_r)+A(u x_r) B(v x_r)) \text{Sin}\left(\frac{x_r}{2}\right)\right)+2 \text{Ci}\left(-\frac{1}{6}
\left(3+\sqrt{3} u+\sqrt{3} v\right) x_r\right) \left((A(u x_r) A(v x_r)-B(u x_r) B(v x_r))
\text{Cos}\left(\frac{x_r}{2}\right)+(A(v x_r) B(u x_r)+A(u x_r) B(v x_r)) \text{Sin}\left(\frac{x_r}{2}\right)\right)+2
\left((A(v x_r) B(u x_r)-A(u x_r) B(v x_r)) \text{Cos}\left(\frac{x_r}{2}\right)-(A(u
x_r) A(v x_r)+B(u x_r) B(v x_r)) \text{Sin}\left(\frac{x_r}{2}\right)\right) \text{Si}\left(\frac{1}{2}
\left(-1-\frac{u}{\sqrt{3}}+\frac{v}{\sqrt{3}}\right) x_r\right)+2 \left((-A(v x_r) B(u x_r)+A(u x_r)
B(v x_r)) \text{Cos}\left(\frac{x_r}{2}\right)-(A(u x_r) A(v x_r)+B(u x_r) B(v
x_r)) \text{Sin}\left(\frac{x_r}{2}\right)\right) \text{Si}\left(\frac{1}{6} \left(-3+\sqrt{3} u-\sqrt{3} v\right) x_r\right)+2
\left((A(v x_r) B(u x_r)+A(u x_r) B(v x_r)) \text{Cos}\left(\frac{x_r}{2}\right)+(A(u
x_r) A(v x_r)-B(u x_r) B(v x_r)) \text{Sin}\left(\frac{x_r}{2}\right)\right) \text{Si}\left(\frac{1}{6}
\left(-3+\sqrt{3} u+\sqrt{3} v\right) x_r\right)+2 \left((A(v x_r) B(u x_r)+A(u x_r) B(v
x_r)) \text{Cos}\left(\frac{x_r}{2}\right)+(-A(u x_r) A(v x_r)+B(u x_r) B(v x_r))
\text{Sin}\left(\frac{x_r}{2}\right)\right) \text{Si}\left(\frac{1}{6} \left(3+\sqrt{3} u+\sqrt{3} v\right) x_r\right)\right)
\end{dmath}
%\end{equation}

\begin{equation}{\label{cis2}}
\begin{aligned}
\mathcal{I}_{s2} =& \frac{1}{36 u^2 v^2} \Biggl( 3 \cos (x_r)  \biggl(u^4 x_r^2-3 u^2  (x_r^2+12 )+ (v^2-3 )  (v^2 x_r^2-36 ) \biggr)\\
  &-x_r \sin (x_r)  \biggl(u^2  (v^2 x_r^2+54 )+54  (v^2-3 ) \biggr) \Biggr)
\end{aligned}
\end{equation}

$\mathcal{I}_{c1}=$
\begin{dmath}{\label{cic1}}
\frac{3}{8 u^3 v^3} \left(-3+u^2+v^2\right)^2 \left(-B(u x_r) \left((3 \pi -2 \mathcal{V}) B(v x_r) \cos\left(\frac{x_r}{2}\right)-(\pi
+2 \mathcal{V}) A(v x_r) \sin\left(\frac{x_r}{2}\right)\right)-(\pi +2 \mathcal{V}) A(u x_r) \left(A(v
x_r) \cos\left(\frac{x_r}{2}\right)-B(v x_r) \sin\left(\frac{x_r}{2}\right)\right)+2 \text{Ci}\left(\frac{1}{6}
\left(-3+\sqrt{3} u+\sqrt{3} v\right) x_r\right) \left(-\left((A(v x_r) B(u x_r)+A(u x_r) B(v
x_r)) \cos\left(\frac{x_r}{2}\right)\right)+(-A(u x_r) A(v x_r)+B(u x_r) B(v
x_r)) \sin\left(\frac{x_r}{2}\right)\right)+2 \text{Ci}\left(-\frac{1}{6} \left(3+\sqrt{3} u+\sqrt{3} v\right) x_r\right)
\left((A(v x_r) B(u x_r)+A(u x_r) B(v x_r)) \cos\left(\frac{x_r}{2}\right)+(-A(u
x_r) A(v x_r)+B(u x_r) B(v x_r)) \sin\left(\frac{x_r}{2}\right)\right)+2 \text{Ci}\left(\frac{1}{6}
\left(-3+\sqrt{3} u-\sqrt{3} v\right) x_r\right) \left((A(v x_r) B(u x_r)-A(u x_r) B(v x_r))
\cos\left(\frac{x_r}{2}\right)+(A(u x_r) A(v x_r)+B(u x_r) B(v x_r)) \sin\left(\frac{x_r}{2}\right)\right)+2
\text{Ci}\left(-\frac{1}{6} \left(3+\sqrt{3} u-\sqrt{3} v\right) x_r\right) \left((-A(v x_r) B(u x_r)+A(u
x_r) B(v x_r)) \cos\left(\frac{x_r}{2}\right)+(A(u x_r) A(v x_r)+B(u x_r)
B(v x_r)) \sin\left(\frac{x_r}{2}\right)\right)+2 \left(-\left((A(u x_r) A(v x_r)+B(u
x_r) B(v x_r)) \cos\left(\frac{x_r}{2}\right)\right)+(-A(v x_r) B(u x_r)+A(u
x_r) B(v x_r)) \sin\left(\frac{x_r}{2}\right)\right) \text{Si}\left(\frac{1}{2} \left(-1-\frac{u}{\sqrt{3}}+\frac{v}{\sqrt{3}}\right)
x_r\right)+2 \left(-\left((A(u x_r) A(v x_r)+B(u x_r) B(v x_r)) \cos\left(\frac{x_r}{2}\right)\right)+(A(v
x_r) B(u x_r)-A(u x_r) B(v x_r)) \sin\left(\frac{x_r}{2}\right)\right) \text{Si}\left(\frac{1}{6}
\left(-3+\sqrt{3} u-\sqrt{3} v\right) x_r\right)+2 \left((A(u x_r) A(v x_r)-B(u x_r) B(v
x_r)) \cos\left(\frac{x_r}{2}\right)-(A(v x_r) B(u x_r)+A(u x_r) B(v x_r))
\sin\left(\frac{x_r}{2}\right)\right) \text{Si}\left(\frac{1}{6} \left(-3+\sqrt{3} u+\sqrt{3} v\right) x_r\right)+2 \left((-A(u
x_r) A(v x_r)+B(u x_r) B(v x_r)) \cos\left(\frac{x_r}{2}\right)-(A(v x_r)
B(u x_r)+A(u x_r) B(v x_r)) \sin\left(\frac{x_r}{2}\right)\right) \text{Si}\left(\frac{1}{6}
\left(3+\sqrt{3} u+\sqrt{3} v\right) x_r\right)\right)
\end{dmath}

\begin{equation}{\label{cic2}}
\begin{aligned}
\mathcal{I}_{c2} &= -\frac{1}{36 u^2 v^2} \Bigl(3\sin (x_r)  (u^4 x_r^2-3 u^2  (x_r^2+12 )+ (v^2-3 )  (v^2 x_r^2-36 ) )\\
 &+x_r \cos (x_r)  (u^2  ( v^2 x_r^2+54 )+54  (v^2-3 ) )\Bigr)
\end{aligned}
\end{equation} 

\begin{figure}
	\begin{center}	
		\includegraphics[scale=0.4]{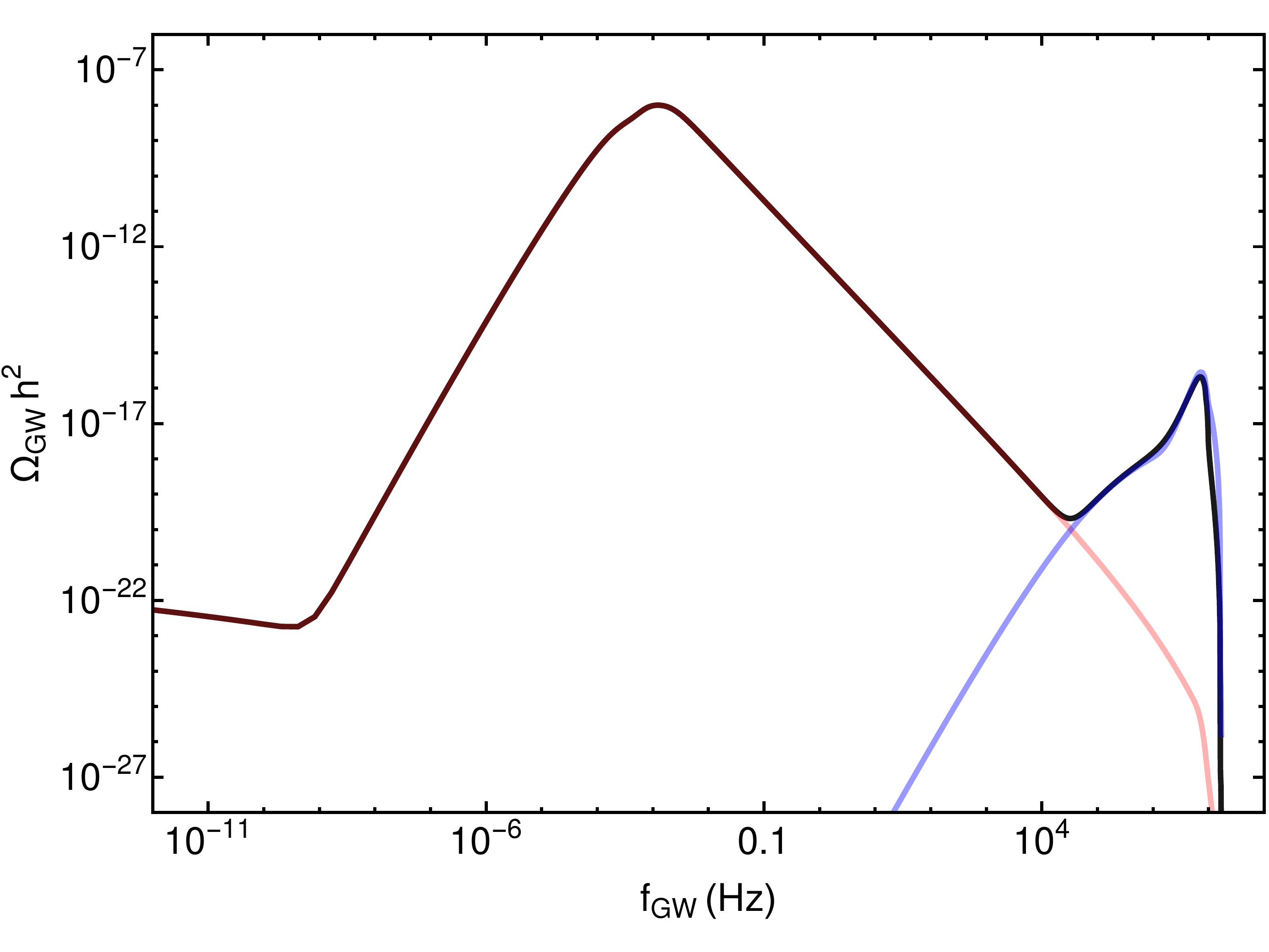}
		\vskip 4pt
		\caption{Comparison of contributions from pure RD approximation (light red) for $k \ll k_r$ limit and from $k \gg k_r$ limit result (light blue) along with the full numerical result (solid black).}
		\label{reeeh}
	\end{center}
\end{figure}

Another way of breaking the terms of $\mathcal{I}_{c} $ and $\mathcal{I}_{s} $ is to identify the terms involving different powers of $x_r$. It is possible to show that, the term involving lowest power of $x_r$ contributes dominantly for small $k$ regime ($k \ll kr$ or $x_r \ll 1 $) and reproduces standard pure RD era formula. This term is also primarily responsible for the first peak in $\Omega_{GW}h^2$, as shown in figure \ref{reeeh}.  The term with highest order of $x_r$, contributes dominantly  for large $k$ regime($k \gg kr$ or $x_r \gg1 $) , and  leads to the second peak in figure \ref{reeeh}. We have also obtained the analytical forms of $\mathcal{I}_{c} $ and $\mathcal{I}_{s} $ for large $k$ regime which are given below.
%\\
%\textbf{$\mathcal{I}_{s}$ in the large $x_r$ limit :}
\vskip 10 pt
\noindent
$\mathcal{I}_{s} (x_r \gg 1) \simeq$
\begin{dmath}{\label{cislxr}}
-\frac{\left(-3+u^2+v^2\right)^2 x_r^4 }{3456 u v} \left(2 \cos\left(\frac{1}{6} \left(3+\sqrt{3} u-\sqrt{3} v\right) x_r\right) \text{Ci}\left(\frac{1}{6}
\left(-3+\sqrt{3} u-\sqrt{3} v\right) x_r\right)+2 \cos\left(\frac{1}{6} \left(-3+\sqrt{3} u-\sqrt{3} v\right) x_r\right) \text{Ci}\left(-\frac{1}{6}
\left(3+\sqrt{3} u-\sqrt{3} v\right) x_r\right)-2 \cos\left(\frac{1}{6} \left(3+\sqrt{3} u+\sqrt{3} v\right) x_r\right) \text{Ci}\left(\frac{1}{6}
\left(-3+\sqrt{3} u+\sqrt{3} v\right) x_r\right)-2 \cos\left(\frac{1}{6} \left(-3+\sqrt{3} u+\sqrt{3} v\right) x_r\right) \text{Ci}\left(-\frac{1}{6}
\left(3+\sqrt{3} u+\sqrt{3} v\right) x_r\right)-\pi  \sin\left(\frac{1}{6} \left(-3+\sqrt{3} u-\sqrt{3} v\right) x_r\right)+\pi
 \sin\left(\frac{1}{6} \left(3+\sqrt{3} u-\sqrt{3} v\right) x_r\right)+\pi  \sin\left(\frac{1}{6} \left(-3+\sqrt{3} u+\sqrt{3}
v\right) x_r\right)+2 \mathcal{V} \sin\left(\frac{1}{6} \left(3+\sqrt{3} u+\sqrt{3} v\right) x_r\right)-2 \sin\left(\frac{1}{6}
\left(-3+\sqrt{3} u-\sqrt{3} v\right) x_r\right) \text{Si}\left(\frac{1}{2} \left(-1-\frac{u}{\sqrt{3}}+\frac{v}{\sqrt{3}}\right)
x_r\right)+2 \sin\left(\frac{1}{6} \left(3+\sqrt{3} u-\sqrt{3} v\right) x_r\right) \text{Si}\left(\frac{1}{6} \left(-3+\sqrt{3}
u-\sqrt{3} v\right) x_r\right)-2 \sin\left(\frac{1}{6} \left(3+\sqrt{3} u+\sqrt{3} v\right) x_r\right) \text{Si}\left(\frac{1}{6}
\left(-3+\sqrt{3} u+\sqrt{3} v\right) x_r\right)-2 \sin\left(\frac{1}{6} \left(-3+\sqrt{3} u+\sqrt{3} v\right) x_r\right) \text{Si}\left(\frac{1}{6}
\left(3+\sqrt{3} u+\sqrt{3} v\right) x_r\right)\right),
\end{dmath}
\vskip 10 pt
%\textbf{$\mathcal{I}_{c}$ in the large $x_r$ limit :}
\noindent

$\mathcal{I}_{c} (x_r \gg 1) \simeq$
\begin{dmath}{\label{ciclxr}}
-\frac{\left(-3+u^2+v^2\right)^2 x_r^4}{3456 u v} \left(\pi  \cos\left(\frac{1}{6} \left(-3+\sqrt{3} u-\sqrt{3} v\right) x_r\right)+\pi
 \cos\left(\frac{1}{6} \left(3+\sqrt{3} u-\sqrt{3} v\right) x_r\right)-\pi  \cos\left(\frac{1}{6} \left(-3+\sqrt{3} u+\sqrt{3}
v\right) x_r\right)+2 \mathcal{V} \cos\left(\frac{1}{6} \left(3+\sqrt{3} u+\sqrt{3} v\right) x_r\right)+2 \text{Ci}\left(-\frac{1}{6}
\left(3+\sqrt{3} u-\sqrt{3} v\right) x_r\right) \sin\left(\frac{1}{6} \left(-3+\sqrt{3} u-\sqrt{3} v\right) x_r\right)-2 \text{Ci}\left(\frac{1}{6}
\left(-3+\sqrt{3} u-\sqrt{3} v\right) x_r\right) \sin\left(\frac{1}{6} \left(3+\sqrt{3} u-\sqrt{3} v\right) x_r\right)-2 \text{Ci}\left(-\frac{1}{6}
\left(3+\sqrt{3} u+\sqrt{3} v\right) x_r\right) \sin\left(\frac{1}{6} \left(-3+\sqrt{3} u+\sqrt{3} v\right) x_r\right)+2 \text{Ci}\left(\frac{1}{6}
\left(-3+\sqrt{3} u+\sqrt{3} v\right) x_r\right) \sin\left(\frac{1}{6} \left(3+\sqrt{3} u+\sqrt{3} v\right) x_r\right)+2 \cos\left(\frac{1}{6}
\left(-3+\sqrt{3} u-\sqrt{3} v\right) x_r\right) \text{Si}\left(\frac{1}{2} \left(-1-\frac{u}{\sqrt{3}}+\frac{v}{\sqrt{3}}\right)
x_r\right)+2 \cos\left(\frac{1}{6} \left(3+\sqrt{3} u-\sqrt{3} v\right) x_r\right) \text{Si}\left(\frac{1}{6} \left(-3+\sqrt{3}
u-\sqrt{3} v\right) x_r\right)-2 \cos\left(\frac{1}{6} \left(3+\sqrt{3} u+\sqrt{3} v\right) x_r\right) \text{Si}\left(\frac{1}{6}
\left(-3+\sqrt{3} u+\sqrt{3} v\right) x_r\right)+2 \cos\left(\frac{1}{6} \left(-3+\sqrt{3} u+\sqrt{3} v\right) x_r\right) \text{Si}\left(\frac{1}{6}
\left(3+\sqrt{3} u+\sqrt{3} v\right) x_r\right)\right).
\end{dmath}

%%%%%%%
%\newpage
\bibliographystyle{JHEP}
\bibliography{bibfl1}
%%%%%%%
\end{document}